\documentclass{article}
\usepackage[english]{babel}

\usepackage{ amssymb }
\usepackage[most]{tcolorbox}

\usepackage[letterpaper,top=2cm,bottom=2cm,left=3cm,right=3cm,marginparwidth=1.75cm]{geometry}

\usepackage[]{babelbib}

\usepackage{cite}

\usepackage{amsmath}
\usepackage{graphicx}
\usepackage[colorlinks=true, allcolors=blue]{hyperref}

\makeatletter
\renewcommand{\fnum@figure}{Fig. \thefigure}
\makeatother

\usepackage{xcolor}
\usepackage{soul}
\sethlcolor{lightgray} 
\usepackage{authblk}
\usepackage{caption}
\usepackage{subcaption}
\newtcolorbox[auto counter]{pabox}[2][]{%
colback=blue!5!white,colframe=blue!75!black,fonttitle=\bfseries,
title=Box ~\thetcbcounter: #2,#1}
\usepackage{cleveref}
\title{Optical Label-Free Microscopy Characterization of Dielectric Nanoparticles}

\author[1,*]{Berenice García Rodríguez}
\author[2,*]{Erik Olsén}
\author[1,*]{Fredrik Skärberg}
\author[1]{Giovanni Volpe}
\author[2]{Fredrik Höök}
\author[1]{Daniel Sundås Midtvedt}

\affil[1]{{\small Department of Physics, University of Gothenburg, Gothenburg, Sweden}}

\affil[2]{{\small Department of Physics, Chalmers University of Technology, Gothenburg, Sweden}}

\affil[*]{{\small These authors contributed equally to this work, and are listed in alphabetical order.}}

\begin{document}
\maketitle

\begin{abstract}
    \noindent In order to relate nanoparticle properties to function, fast and detailed particle characterization, is needed. The ability to characterize nanoparticle samples using optical microscopy techniques has drastically improved over the past few decades; consequently, there are now numerous microscopy methods available for detailed characterization of particles with nanometric size. However, there is currently no ``one size fits all'' solution to the problem of nanoparticle characterization. Instead, since the available techniques have different detection limits and deliver related but different quantitative information, the measurement and analysis approaches need to be selected and adapted for the sample at hand.
    In this tutorial, we review the optical theory of single particle scattering and how it relates to the differences and similarities in the quantitative particle information obtained from commonly used microscopy techniques, with an emphasis on nanometric (submicron) sized dielectric particles. 
    Particular emphasis is placed on how the optical signal relates to mass, size, structure, and material properties of the detected particles and to its combination with diffusivity-based particle sizing.
    We also discuss emerging opportunities in the wake of new technology development, with the ambition to guide the choice of measurement strategy based on various challenges related to different types of nanoparticle samples and associated analytical demands.
\end{abstract}

\section*{Key learning points}
After reading this tutorial, the reader will have the tools to:
\begin{enumerate}
    \item Describe how an image of a nanoparticle is formed in a light scattering microscope.
    \item Relate scattering microscopy measurements to particle properties.
    \item Identify the most suitable scattering microscopy techniques for specific samples and scientific questions.
    \item Perform nanoparticle characterization using state-of-the-art analysis techniques, provided by ready-to-use and adaptable Python notebooks.
\end{enumerate}

\section{Introduction}

Accurate nanoparticle characterization in terms of size, shape, and composition in complex biological environments is crucial to understanding the relation between nanoparticle structure and function as well as to achieving the full potential of nanoparticle-assisted applications within several fields, including drug delivery\cite{jindal2017effect,chandrakala2022review} and medical diagnostics\cite{chen2016nanochemistry,oliveira2023engineering}.
Traditionally, such characterization has been performed on the individual particle level using high spatial resolution methods, such as cryogenic transmission electron microscopy (cryo-TEM)\cite{luque2020cryo}, whereas light scattering techniques have been employed to perform quick nanoparticle characterization on an ensemble level. 
Two examples of such light scattering techniques routinely used to characterize nanoparticle suspensions are dynamic light scattering (DLS)\cite{stetefeld2016dynamic}, connecting particle diffusivity to size, and multi-angle light scattering (MALS) to characterize both size and structure\cite{wyatt1993light,wyatt1998submicrometer}. 

However, none of these approaches are satisfactory to characterize heterogeneous nanoparticle samples: electron microscopy is an \textit{ex-situ} approach suffering from low throughput, while ensemble approaches measure an averaged signal over many individual particles, masking their underlying heterogeneity\cite{chan2023role}. This is particularly problematic for biological nanoparticles, which often display a pronounced heterogeneity in terms of size and composition, which furthermore may be a deciding factor for their biological function\cite{rabanel2019nanoparticle}.

In this context, single particle characterization using label-free light scattering microscopy has emerged as an alternative route, achieving widespread use in the last two decades. In fact, the first use of optical microscopy for the characterization of nanoparticles was done over 100 years ago, which relied on orthogonal illumination and detection pathways to achieve darkfield microscopy\cite{siedentopf1902uber}. Although nanoparticles are smaller than the optical resolution limit, it is still possible to detect individual nanoparticles as long as the signal-to-noise is high enough, which in turn enables detailed measurements on the single particle level. Modern implementations image the particles onto sensitive cameras and quantify a combination of the scattering signal and particle motion to achieve high-throughput characterization of particle samples\cite{priest2021scattering,van2014refractive,midtvedt2019size,kashkanova2021precision,nanochannel2022}. From the nanoparticle motion, the size is estimated \textit{via} the Stokes-Einstein relation, which relates diffusivity to size for spherical particles in a viscous medium\cite{bian2016111}. The use of nanoparticle motion to estimate size has achieved widespread application under the name Nanoparticle Tracking Analysis (NTA) and exemplifies how scattering microscopy extends traditional ensemble-based characterization approaches to characterize nanoparticles with single nanoparticle resolution\cite{filipe2010critical,midtvedt2019size,kashkanova2021precision}.

Going beyond diffusivity-based particle sizing, over the past decade numerous optical microscopy methods have been developed, aiming at multiparametric nanoparticle characterization with single-particle resolution\cite{van2014refractive,kashkanova2021precision,midtvedt2019size,midtvedt2021fast,ortiz2022simultaneous,nanochannel2022,nguyen2023label,altman2023machine,olsen2024dual}. All these techniques rely on the following fundamental observation: the amount of light scattered and absorbed by an object is to a first approximation proportional to its volume and refractive index contrast relative to the surrounding medium. The refractive index is a complex-valued material-specific property dictating the efficiency of a material to scatter and absorb light\cite{bohren2008absorption}. The real part of the refractive index governs light scattering, while the imaginary part of the refractive index governs light absorption. Thus, optical nanoparticle characterization can distinguish between different types of particles based on their ability to scatter and absorb light\cite{adhikari2020photothermal,priest2021scattering}. In the specific case of biological nanoparticles, the total amount of light scattered is to a first approximation proportional to particle mass \cite{zangle2014live,doi:10.1126/science.aar5839,kowal2024electrophoretic}. In addition to the scattering amplitude, the relative amount of scattering to different scattering angles can also be used to characterize nanoparticle samples\cite{fu2013directional,olsen2024dual}.

\begin{figure}[!ht]
\centering
\includegraphics[width=0.9\linewidth]{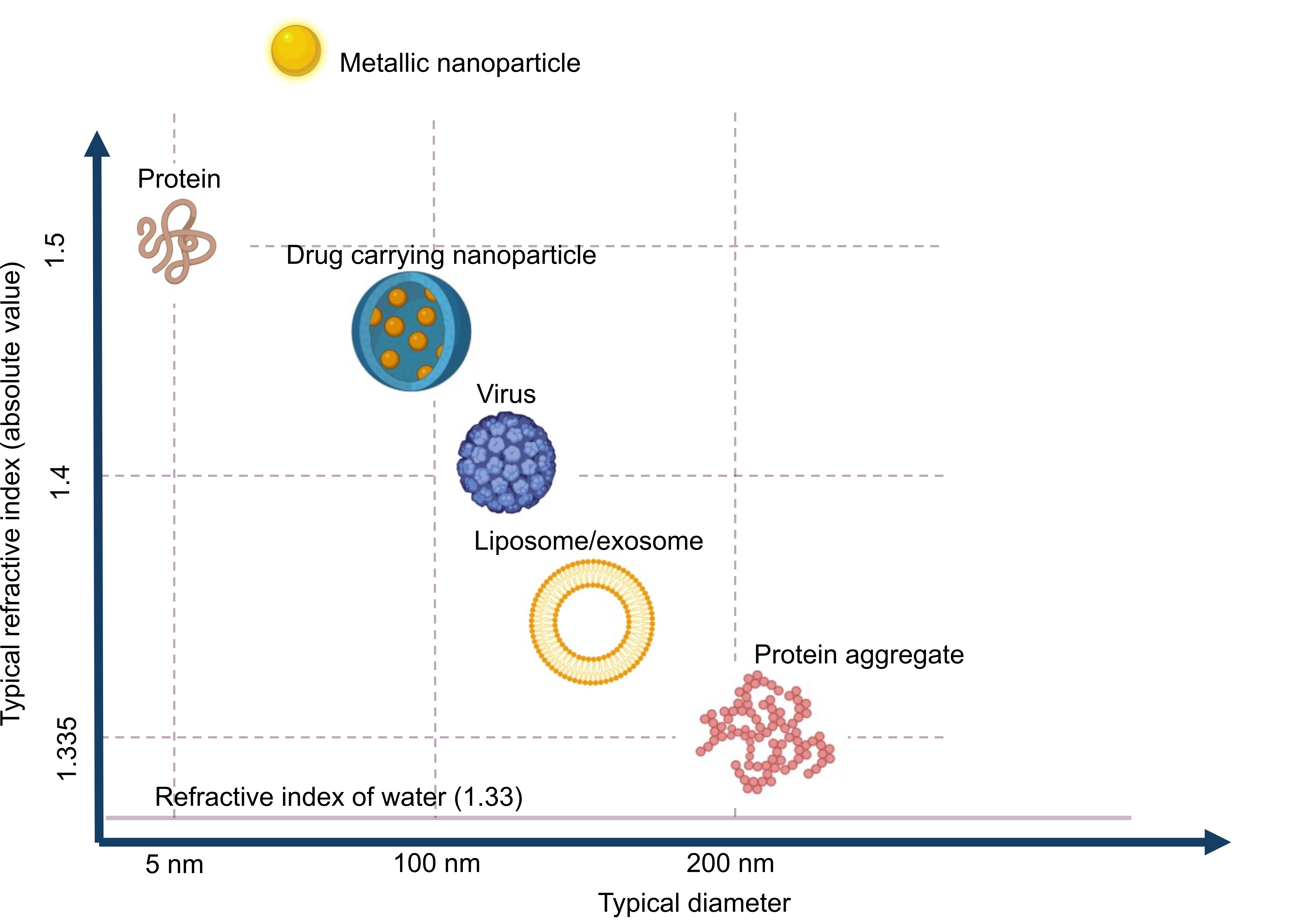}
\caption{\textbf{Typical nanoparticles studied in scattering microscopy, organized by size and refractive index.} The size range of nanoparticles typically studied by scattering microscopy techniques ranges from individual biomolecules to large biomolecular complexes. }
\label{overview}
\end{figure}

To give some specific examples, some types of nanoparticles that can be characterized using scattering microscopy are highlighted in Figure \ref{overview}, showcasing typical values of the respective sizes and refractive indices. For objects much smaller than the wavelength of the illuminating light, such as individual biomolecules, the relative amount of light scattering at different angles has a weak particle size dependence\cite{wyatt1993light,wyatt1998submicrometer}. For such particles, it is sufficient to determine the scattering amplitude in a limited range of angles to characterize particle mass. This has been utilized to determine the mass of individual proteins\cite{young2018quantitative, priest2021scattering, doi:10.1126/science.aar5839}. For larger biomolecular complexes, such as viruses, liposomes, and protein aggregates, the scattered light amplitude integrated over all scattering angles is still related to their mass. However, interference effects between the scattered light from individual molecular elements within the complexes generate a directionality of the scattered light, such that the light amplitude measured in a scattering microscope will depend on the measurement geometry\cite{bohren2008absorption}. For instance, a liposome, consisting of a lipid bilayer shell with a water-filled core, will scatter light differently from a drug-containing nanoparticle of the same size and mass, simply due to a different spatial distribution of biomolecules. This fact has been utilized to distinguish empty liposomes from exosomes filled with biological material through simultaneous characterization of scattering amplitude and size \cite{kashkanova2021precision}.

Metallic nanoparticles present another example of widely used nanoparticles that can be characterized using scattering microscopy\cite{priest2021scattering}. Such nanoparticles interact much more strongly with the illuminating light compared to biological nanoparticles of the same particle size due to the refractive index difference between gold and water is larger than that of biomolecules and water (Figure \ref{overview}). Moreover, in contrast to biological nanoparticles, they typically display considerable light absorption in addition to light scattering, as a result of plasmonic resonance. This plasmonic resonance, in turn, depends on particle size and shape\cite{maier2007plasmonics}. Thus, to fully characterize such nanoparticles it is important to use an experimental design capable of quantifying both the scattered as well as the absorbed light.

From the above considerations, it becomes clear that while individual biomolecules can be characterized based on the measured scattered light alone, the characterization of more complex structures requires an experimental design and data analysis pipeline that is optimized for the specifics of the sample.
Therefore, although method development and refinement are likely to continue and further expand the information that can be extracted from microscopy images, it will always be imperative to choose an experimental design that maximizes the amount of useful information about the investigated sample in the recorded scattering pattern, and an image analysis approach that optimally utilizes that information. The purpose of this tutorial is to guide the reader in this process by providing a \textbf{1} a theoretical understanding of how the image formed in a light scattering microscope, \textbf{2} an understanding of how to quantify physical properties of the measured nanoparticles from their corresponding microscopy image, \textbf{3} clear guidelines on how to choose the right measurement modality for specific purposes, and \textbf{4} a toolbox for performing optical nanoparticle characterization using scattering microscopy, in the form of Python notebooks containing ready-to-use code for particle detection and characterization. The importance of considering these aspects is highlighted by a few examples from the literature.



\section{Image formation in a scattering microscope}

\begin{figure}[!ht]
\centering
\includegraphics[width=0.9\linewidth]{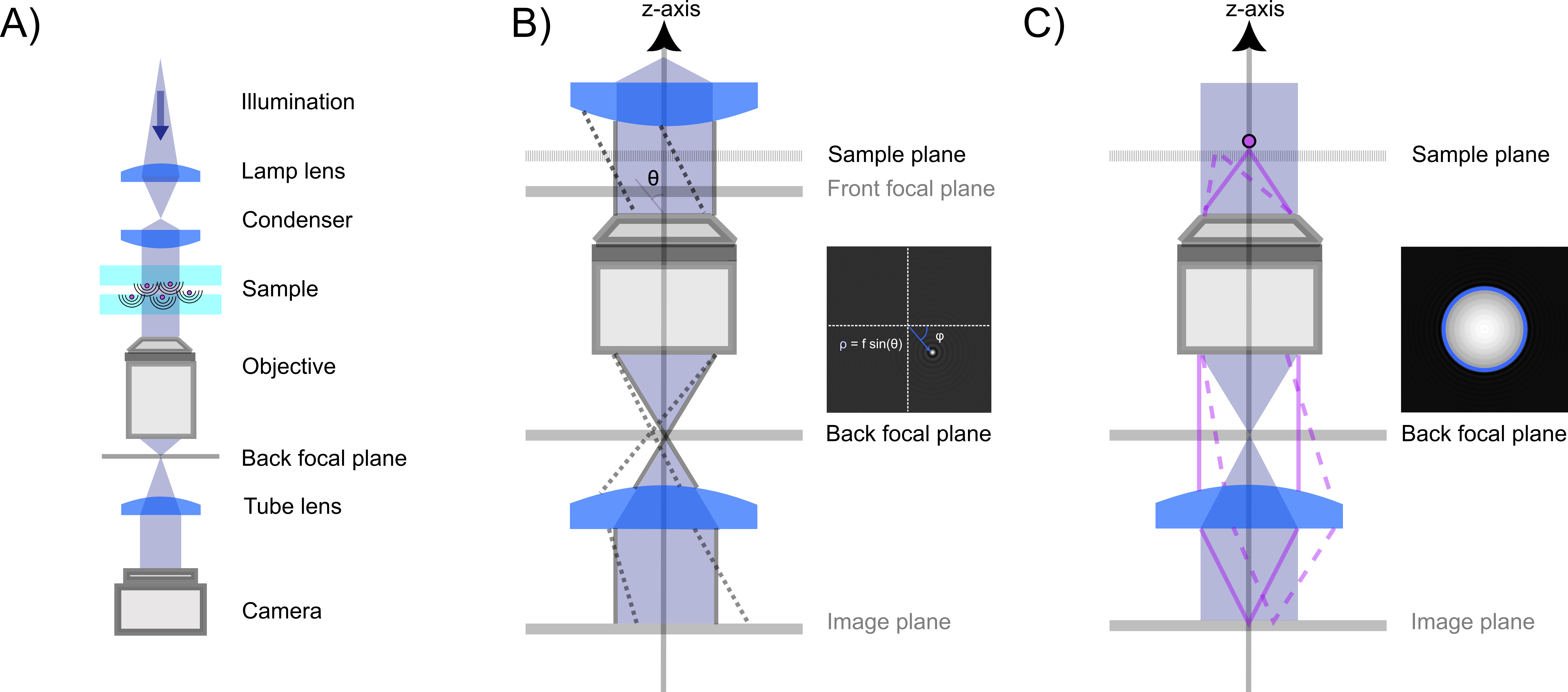}
\caption{\textbf{Propagation of the scattered light through a scattering microscope.} A) A general scattering microscopy setup. The sample is illuminated by a plane wave. Both the illuminating light and the scattered light from particles in the sample propagate through the optical system and are recorded by a camera. B) The image obtained by a scattering microscope is largely determined by the objective lens. Plane waves entering into the objective are transformed into a focused spot at the back focal plane. When the objective is illuminated by a plane wave impinging on the objective at an angle $\theta$ with respect to the optical axis and at an angle $\phi$ relative to the horizontal axis, this tightly focused spot is offset a distance $\rho=f\sin\theta$ away from the optical axis. C) The scattered field from a nanoparticle can be considered a linear combination of many plane waves, incident on the objective with different angles $\theta$ and $\phi$. Each such plane wave produces a similar spot as described in B). Summing up, all these plane wave contributions produce a field at the back focal plane as shown in the inset, which, in the case of particles much smaller than the wavelength of light to a first approximation, reassembles that of a plane wave. Notice the sharp cutoff, highlighted as a circle in the inset. This is due to the limited angular range admitted by the objective and is set by the objective NA.}
\label{backFocalPlane}
\end{figure}

A light scattering microscope is a device where the scattered light from micro- and nanosized objects is collected and recorded by, for example, a camera or photomultiplier tube (Figure \ref{backFocalPlane}A). Most commonly, a microscope objective is used to collect the scattered light from the objects (although lens-free solutions exist as well\cite{ozcan2016lensless}). On the opposite side of the objective, another lens called the tube lens, collects the light from the objective to form an image onto the camera. The objective, tube lens, and camera together define the \textit{optical axis}. 

The properties of the objective and the sample illumination largely determine the scattering information that propagates to the camera in a scattering microscope.
First, consider an objective illuminated by a plane wave, which corresponds to a collimated sample illumination with a constant intensity profile propagating along the optical axis. At the back focal plane of the objective, this wave is focused to a spot centered on the optical axis (Figure \ref{backFocalPlane}B). A plane wave that is tilted by an angle $\theta$ relative to the optical axis will produce a focused spot at the back focal plane that is offset by a distance $f\sin\theta$ relative to the optical axis, where $f$ is the back focal length of the objective (Box \ref{box:objective}). 

\begin{pabox}[label=box:objective]{Focusing of a plane wave by an objective}

From a mathematical standpoint, the microscope objective can be represented by a pupil function $P(\theta)$, defining the transmittance of plane waves with incident angle $\theta$ on the objective (Figure \ref{backFocalPlane}C). For an ideal objective, the pupil function is
\begin{equation}
    P(\theta) = \begin{cases}
        1 & \theta<\theta_{\rm max}\\
        0 & \text{otherwise}
    \end{cases},
\end{equation}
where $\theta_{\rm max}$ is the largest incident angle accepted by the objective. This is related to the numerical aperture (${\rm NA}$) of the objective as ${\rm NA}=n_{\rm NA}\sin\theta_{\rm max}$, where $n_{\rm NA}$ is the refractive index of the media between the objective front lens and the sample.

\hspace*{0.4cm} Now, consider an objective illuminated by a plane wave propagating at an angle $\theta$ relative to the optical axis and at an angle $\phi$ relative to the x-axis. The optical field\footnote{In optical microscopy, the optical (or light) field is often used instead of the electric field. The difference between the two fields is a normalization constant that affects how the fields relate to the measured light intensity, where the measured light intensity at a camera is equal to the squared modulus of the optical field.} at the back focal plane is then given by
\begin{equation}
    E_{\rm bfp}(x,y) = 
    A \hat{P}(k(x-f\sin\theta\cos\phi),k(y-f\sin\theta\sin\phi)),  
\end{equation}
where the function $\hat{P}$ is the transfer function of the objective\footnote{The transfer function is the Fourier transform of the pupil function.}, $z=f$ is the back-focal length of the objective lens, and $k=2\pi/\lambda$ where $\lambda$ is the wavelength of light (Figure \ref{backFocalPlane}B). For an objective without optical aberrations, the transfer function under the paraxial approximation is given by\cite{chandler2019spatio}
\begin{equation}
\hat{P}(p,q)=2\frac{J_{1}({\rm NA}\sqrt{p^2+q^2})}{{\rm NA}\sqrt{p^2+q^2}},
\end{equation}
where $J_{1}$ is the first-order Bessel function. Thus, all incoming plane waves are transformed into tightly focused spots in the back-focal plane, and the position of each spot in the back-focal plane depends on its incident angle. 
\end{pabox}

As it turns out, a general incident optical field can be described as a linear combination of plane waves with different incident angles $\theta$ relative to the optical axis (Box \ref{box:decomposition}), and $\phi$ relative to an at this point arbitrarily defined $x$-axis perpendicular to the optical axis\cite{goodman2005introduction}.
Connecting this to the specific topic of this tutorial, consider a nanoparticle located at the front focal plane of the objective (Figure \ref{backFocalPlane}C), illuminated by a plane wave propagating along the optical axis. This produces a scattered optical field $E$, which can be decomposed as a sum of plane waves $\hat{E}(\theta,\phi)$, each propagating with specific angles $\theta$ and $\phi$. 
The field at the back focal plane can then be obtained by summing the contributions from all such plane-wave components, each of which produces a focused spot as in Figure \ref{backFocalPlane}B. Although the general expression is fairly complicated, an approximate form of the field at the back-focal plane can be obtained by utilizing that each plane wave component contributes to the summation only close to the center of the corresponding focused spot. Under this approximation, the field at the back focal plane is given by
\begin{equation}
    E_{\rm bfp}(\rho=f\sin\theta) \approx \hat{E}(\theta,\phi)\;\; \theta<\theta_{\rm max}.
    \label{eq:bfpplanewaves}
\end{equation}
Thus, the field at the back focal plane mirrors the angular distribution of the scattered light, for scattering angles $\theta<\theta_{\rm max}$ (Figure \ref{backFocalPlane}C).

\clearpage
\begin{pabox}[label=box:decomposition]{Decomposition of the optical field in plane wave components}
Similar to how a time-dependent signal can be decomposed to a sum of contributions with different amplitudes and frequencies, an optical field can be decomposed to a sum of plane waves with different wave vectors (or spatial frequencies). Each such plane wave component can mathematically be written as 
\begin{equation}
    \Phi_{\bf k}({\bf x}) = e^{i {\bf k\cdot x}}.
\end{equation}
Consider a monochromatic optical field $E$ (meaning that the field consists of light with a single frequency), evaluated on a plane (the $xy$-plane) perpendicular to the optical axis. Its plane wave decomposition reads
\begin{equation}
    E({\bf x}) = \int e^{i {\bf k\cdot x}} \hat{E}({\bf k}) d{\bf k},
    \label{eq:transf1}
\end{equation}
where ${\bf k}$ is the projection of the wave vector of the plane wave component $\Phi_{\bf k}$ on the $xy$-plane, and $\hat{E}({\bf k})$ is a coefficient describing both amplitude and phase of the plane wave component $\Phi$. These components are found through the inverse transform of Eq. \eqref{eq:transf1},
\begin{equation}
    \hat{E}({\bf k}) = (2\pi)^{-2} \int e^{-i {\bf k\cdot x}} E({\bf x}) d{\bf x}.
    \label{eq:transf2}
\end{equation}

\hspace*{0.4cm} Since $\mathbf{k}$ is a projection of the wave vector onto the $xy$-plane orthogonal to the optical axis, it follows that $|\mathbf{k}|=k\sin\theta$, where $k=(2\pi n_{\rm m}/\lambda)$ is the wave number of the light and $\theta$ is the angle between the propagation direction of the plane wave component and the optical axis. The mathematically versed reader may recognize that these equations resemble Fourier transforms. In fact, the plane wave decomposition is identical to the Fourier transform, except that the integration over $\mathbf{k}$ is restricted for the reason outlined above. Specifically, denoting the Fourier transform operator by $\mathcal{F}$, the above equations can be written as
\begin{align}
    E(\mathbf{x}) = & \mathcal{F}^{-1}[\hat{E}(\mathbf{k}) C(\mathbf{k})] \\
    \hat{E}(\mathbf{k}) = &  \mathcal{F}[E(\mathbf{k})], 
\end{align}
where $C(\mathbf{k}) = 1$ for $|\mathbf{k}|<k$, and $0$ otherwise.
In the case of a circularly symmetric scattered field, as is often the case for the scattered field from nanoparticles, the decompositions above are more conveniently expressed in polar coordinates. In this case, one obtains
\begin{equation}
    E(\rho) = \int \sin\theta\cos\theta \hat{E}(\theta) J_0(k\rho\sin\theta) d\theta, 
    \label{eq:iFTransform}
\end{equation}
where the integration now runs over the propagation angle $\theta$ with respect to the optical axis instead of the wave vector projection $\mathbf{k}$. The coefficients $\hat{E}(\theta)$ are similarly found by 
\begin{equation}
    \hat{E}(\theta) = k^2 \int \rho E(\rho)J_0(k\rho\sin\theta) d\rho. 
    \label{eq:FTransform}
\end{equation}

\end{pabox}

To form an image of the sample on the camera, a lens (tube lens) is placed between the objective and camera, such that an incoming plane wave to the objective enters the camera as a plane wave, and light scattered from a point source at the focal plane of the objective forms a focused image on the camera. As a result of this, the plane wave components of the field at the camera plane reproduce exactly the plane wave components impinging on the objective, for all $\theta<\theta_{\rm max}$. Mathematically, this is expressed as
\begin{equation}
    \hat{E}_{\rm cam}(\theta,\phi) = \hat{E}(\theta,\phi)P(\theta).
\end{equation}

Now, we have a mathematical framework for relating the field at the focal plane to the fields at the back-focal plane and the camera. This framework involves decomposing the field into its plane wave components, applying the optical transfer function, and summing up the contributions from the individual components.

In the context of scattering microscopes, the optical field at the focal plane is typically a superposition of the illuminating field $E_{\rm ill}$, and the field scattered by objects in the sample $E_{\rm sca}$. The field at the camera is then given by 
\begin{equation}
    \hat{E}_{\rm camera}(\theta,\phi)= \hat{E}_{\rm ill}(\theta)P(\theta) + \hat{E}_{\rm sca}(\theta,\phi)P(\theta)
\end{equation}
However, the microscope camera does not record the angular components of the incident field; instead, it records the spatial distribution of the incident field intensity. This is given by the modulus squared of the spatial distribution of the optical field. The intensity recorded by the camera can be written as

\begin{equation} 
I_{\rm cam}(x,y) = |E_{\rm ill}+E_{\rm sca}(x,y)|^2= |E_{\rm ill}|^2P(\theta_{\rm ill})+|E_{\rm sca}(x,y)|^2+2|E_{\rm ill}|\Re (E_{\rm sca}(x,y)^*e^{i\varphi})P(\theta_{\rm ill}),
    \label{Eq:intscat}
\end{equation}
where the spatial distributions of the illuminated and scattered fields are related to their plane wave components through Eq \eqref{eq:iFTransform} (Box \ref{box:decomposition}), and $\varphi$ represents the phase difference of the illuminating field relative to the scattered field. For completeness, in Box \ref{box:camint} the contributions from the individual plane wave components to the measured intensity is described.

\begin{pabox}[label=box:camint]{Decomposition of the recorded intensity into plane wave components}
The formalism introduced in Box 1 allows us now to investigate how the plane wave components of the scattered and illuminating light contribute to the recorded intensity. At the camera plane, the scattered field is given by the Fourier transform of the field at the back focal plane of the objective. Since the objective only admits plane wave components with angles relative to the optical axis $\theta\leq\theta_{\rm max}$, it follows that it only admits plane waves with projected wave vectors $|\mathbf{k}|\leq k\sin\theta_{\rm max}$. The scattered field at the camera plane is therefore given by
\begin{equation}
    E_{\rm sca}(\mathbf{x})=\int e^{i {\bf k\cdot x}} \hat{E}_{\rm sca}({\bf k}) \tilde{P}(\mathbf{k}) d{\bf k},
\end{equation}
where $\tilde{P}$ is the pupil function expressed in terms of the projected wave vectors instead of the propagation angle, given by $\tilde{P}(\mathbf{k})=1$ for $|\mathbf{k}|\leq 
 k\sin\theta_{\rm max}$ and $0$ otherwise. In the specific case of a circularly symmetric scattered field, one has 
\begin{equation}
    E_{\rm sca}(\rho) = \int_{0}^{\theta_{\rm max}} \cos\theta \sin\theta\hat{E}_{\rm sca}(\theta) J_0(k\rho\sin\theta) d\theta.
\end{equation}
In this case, one finds for the second and last terms in Eq. \eqref{Eq:intscat}
\begin{align}
|E_{\rm sca}(x,y)|^2 = &\left|\int_{0}^{\theta_{\rm max}} \cos\theta \sin\theta\hat{E}_{\rm sca}(\theta) J_0(k\rho\sin\theta) d\theta\right|^2 \\
|E_{\rm ill}|\Re (E_{\rm sca}(x,y)^*e^{i\varphi})P(\theta_{\rm ill}) =& P(\theta_{\rm ill})|E_{\rm ill}|P(\theta_{\rm ill})\int_{0}^{\theta_{\rm max}} \cos\theta \sin\theta J_0(k\rho\sin\theta) \Re{(\hat{E}_{\rm sca}(\theta)^*e^{i\varphi}})  d\theta.
\end{align}
In this expression, it is assumed that the field reaching the camera is not magnified by the objective. The effect of magnification is essentially to reduce the propagation angle $\theta$ of an optical field with respect to the optical axis. The effect of magnification can be taken into account by replacing the argument of $J_0$ with $Mk\rho\sin\theta$, where $M$ is the magnification. Similarly, the function $\tilde{P}=1$ for $|\mathbf{k}|\leq k\sin\theta_{\rm max}/M$ and $0$ otherwise in the case of a non-unitary magnification.

\end{pabox}






The first two terms of Eq. \eqref{Eq:intscat} describe the intensities of the illuminating field and the scattered field independently. The third term describes the interference of the scattered field with the illuminating field and contains information about the relative phases of the two fields. Notice that the first and last of these terms are relevant only for illumination angles for which $P(\theta_{\rm ill})>0$.

The amount of light scattered from a nanoparticle is generally much smaller than the amplitude of the light that is incident on it. Thus, in a transmission scattering measurement, as depicted in Figure \ref{backFocalPlane}A, the first term of Eq. \eqref{Eq:intscat} will dominate over the last two terms, resulting in a small signal-to-background ratio. 
Three strategies are traditionally used to overcome this signal-to-background ratio limitation for quantitative characterization of subwavelength-sized particles; i) darkfield microscopy enhances the contrast by only allowing the scattered light to reach the camera, ii) interferometric scattering techniques instead focus on quantifying the interferometric term, while 
iii) quantitative field microscopy uses the interference of the field at the camera plane with another optical field to quantify the complex-valued field itself instead of only the real part of the interference between the optical fields of the particle and the background.
All mentioned strategies have in common that they are designed to manipulate one or more of the terms in Eq. \eqref{Eq:intscat} to enhance their performance, and the details of how this is achieved using the three techniques mentioned above are discussed in the following subsections.

\subsection{Darkfield microscopy}
\begin{figure}[!ht]
\centering
\includegraphics[width=1\linewidth]{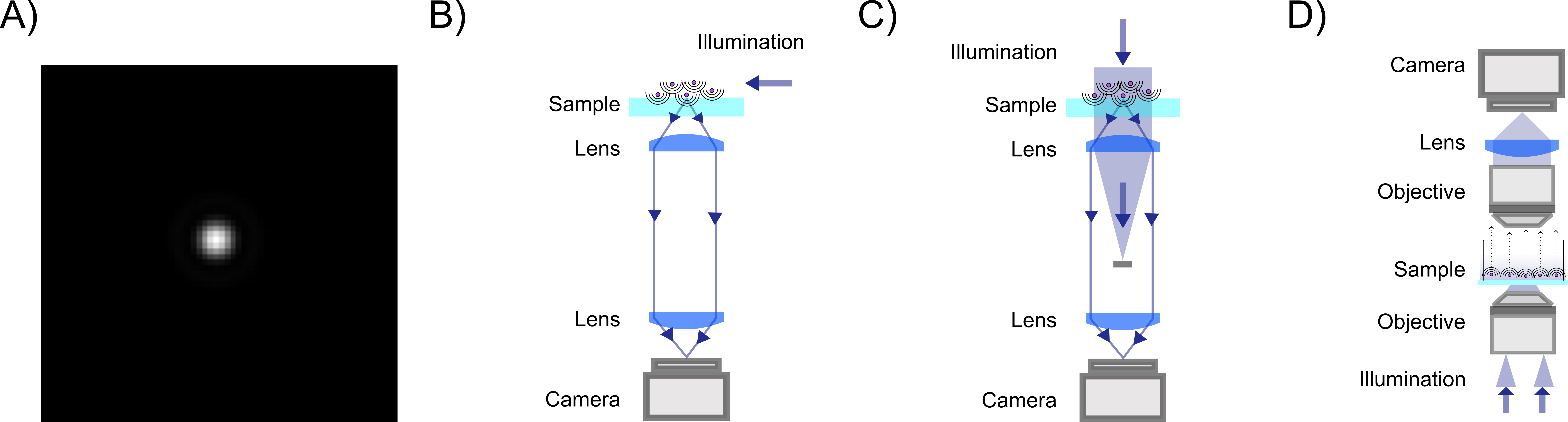}
\caption{\textbf{Darkfield microscopy setups:} A) In darkfield microscopy, the particles are visible on the camera as bright spots against a dark background. B) Darkfield through oblique illumination: the sample is illuminated by light at an angle larger than the maximum angle admitted by the objective so that the illuminating light is prevented from propagating through the optical system. C) Darkfield using spatial blocking: the illumination light is blocked at the back focal plane using a physical filter. D) Evanescent field imaging: one utilizes total internal reflection at the interface between glass and sample to produce an evanescent field. This evanescent field amplitude decays exponentially away from the surface and thus will not reach the camera while scattering from particles close to the interface can be recorded on a camera. }
\label{dark_field_filter}
\end{figure}

Darkfield microscopy is a widely used technique to enhance the contrast of small particles, being a relatively simple but powerful configuration. It has been used for over 100 years\cite{siedentopf1902uber} and is still one of the standard techniques to characterize the hydrodynamical radius of nanoparticles using NTA\cite{van2014refractive,kim2019validation}. Darkfield techniques aim to only allow the scattered light to reach the camera, thereby suppressing all terms except the second term in Eq. \eqref{Eq:intscat}. In this way, the scattered light from the particles appears as bright dots against a dark background (Figure \ref{dark_field_filter}A) (hence the name ``darkfield''). Since there is no background signal, the darkfield signal is given by
\begin{equation}
    I_{\rm DF}=\left|E_{\rm sca}\right|^2 .
\end{equation}
Such background suppression can be achieved by a vast range of microscope configurations, where the choice of configuration affects the relation between the properties of the particles and the measured signal.
The most common ways of achieving such background rejection rely on using an illumination angle outside the range captured by the objective (Figure \ref{dark_field_filter}B), spatial blocking of the excitation beam (Figure \ref{dark_field_filter}C), or an evanescent illumination (Figure \ref{dark_field_filter}D)\cite{priest2021scattering}.  


Using an illumination angle that lies outside the range captured by the objective achieves full background suppression while still allowing for the particle scattering to be measured for a wide range of different suspended particles (Figure \ref{dark_field_filter}B)\cite{siedentopf1902uber,dragovic2011sizing,ortiz2020precise}. This approach is commonly used in NTA setups, in which suspended nanoparticles are tracked and characterized based on their Brownian motion\cite{dragovic2011sizing}. However, the largest angle admitted by the objective must be smaller than the illumination angle for the incoming light not to be collected, which puts a limit on the plane wave components reaching the camera and, therefore, the information content of the microscope images.


Darkfield microscopy can also be achieved through spatial blocking of the excitation beam using an illumination angle within the numerical aperture of the objective (Figure \ref{dark_field_filter}C). It relies on the observation that at the back focal plane of the objective, the illuminating plane wave is tightly focused, while the scattered field from a point source propagates as a plane wave (Figure \ref{backFocalPlane})\cite{goodman2005introduction}. As a result of this, the excitation beam can be blocked from reaching the camera by placing a small non-transmitting filter at the back focal plane\cite{sowa2010simple,weigel2014dark}. 
Specifically, the effect of the filter on the scattered field from the particles can be neglected as long as the physical size of the filter is much smaller than the extent of the scattered field at the back focal plane. This type of darkfield microscope is sensitive to stray light that is not blocked by the optical filter, which limits the background rejection.


Finally, darkfield microscopy can be achieved by using an evanescent field to illuminate the sample (Figure \ref{dark_field_filter}D). An evanescent field is an optical near-field that is present at the interface between two media when the incident light undergoes total internal reflection at the interface. This is the principle of operation of total internal reflection microscopy and waveguide microscopy\cite{ueno2010simple,zhang2022evanescent,sjoberg2021time}. The evanescent field is present only very close to the interface (within a fraction of the wavelength of the illuminating light) and does not propagate to the camera. The light that is scattered from the evanescent field, however, does propagate and can therefore be imaged.
The evanescent field, therefore, limits the imaging to particles close to an interface.


\subsection{Interferometric scattering techniques}
A drawback of the darkfield techniques discussed in the previous section is that the illuminating light is not recorded. Since the scattered field from a nanoparticle is proportional to the illuminating field (see Section \ref{sec:relation} for details), quantitative nanoparticle characterization using darkfield techniques requires either detailed calibration or separate quantification of the light intensity at the sample plane. 

\begin{figure}[!ht]
\centering
\includegraphics[width=0.7\linewidth]{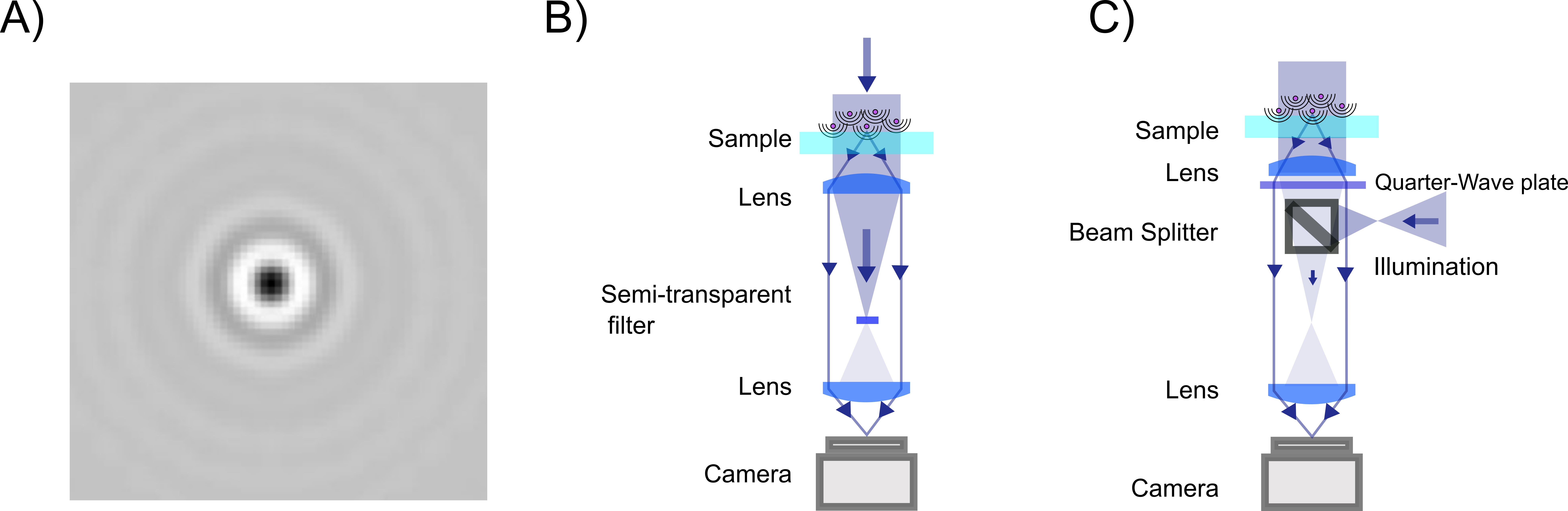}
\caption{\textbf{Interferometric scattering techniques setups:} A) In interferometric scattering microscopy, the scattering patterns are often seen as dark spots against a bright background\cite{cole2017label,doi:10.1126/science.aar5839}, where both the sign and contrast of the signal depends on the relative phase difference between the scattering signal and the background signal (Figure \ref{fig:iscatintensity}A). B) Coherent Brightfield microscopy (COBRI): the scattered light is recorded in transmission, and the relative strength of the illumination light and the scattered light reaching the camera is tuned using a semi-transparent filter. C) Interferometric scattering microscopy (iSCAT): the sample is illuminated from below. Light is partially reflected at the interface between the coverslip and the sample. This partially reflected light is propagated through the optical system and is recorded on the camera. Most of the light is, however, not reflected at the interface and instead produces scattering from nanoparticles in the sample. The scattered field from the nanoparticles interferes with the reflected field to produce interferometric scattering patterns on the camera. }
\label{fig:cobri}
\end{figure}

Interferometric is an approach that circumvents this problem by focusing on quantifying the last term of Eq.~\eqref{Eq:intscat}, explicitly containing the illuminating field $E_{\rm ill}$. The signal is the interference between the particle signal and a background signal, and therefore, the background signal is non-zero, representing the illuminating field amplitude (Figure \ref{fig:cobri}A). This ensures an internal reference to which the scattered field can be compared.

However, given the limited dynamic range of the detector recording the image, the signal-to-background ratio will limit the ability to detect weakly scattering particles in this measurement strategy since the particle contrast can not be enhanced by increasing the illuminating light intensity. One way to overcome this limitation is by introducing a partially transmissive filter centered at the back focal plane of the objective (Figure \ref{fig:cobri}B)\cite{goto2015three,cole2017label,cheng2019high}. The presence of the filter can be represented by a function $T(\theta)$,
\begin{equation}
    T(\theta) = \begin{cases} \epsilon e^{i\varphi_{\rm filt}} & \text{if}~\theta <\theta_{\rm filt}\\
    1 & \text{otherwise}
    \end{cases},
\end{equation}
with $\epsilon<1$, such that the filter attenuates and phase shifts plane waves with incident angles $\theta<\theta_{\rm filt}$\cite{goto2015three,olsen2023label,olsen2024dual}. If $\theta_{\rm filt}\ll \theta_{\rm max}$ and positioned in the back focal plane such that it attenuates the illuminating light before it can reach the camera, the filter will attenuate the illuminating light while leaving the scattered light unaffected.  The recorded intensity is then, to the lowest order in the scattered field
\begin{equation}
    I_{\rm cam} = \epsilon^2 |E_{\rm ill}|^2 + 2\epsilon|E_{\rm ill}|\Re{(e^{i(\varphi_{\rm filt}+\varphi)}E_{\rm sca}^*)}.
    \label{eq:iscat}
\end{equation}
Notice that the first term is quadratic in the attenuation $\epsilon$, while the second term is linear in $\epsilon$. Thus, the relative importance of the two terms can be adjusted by adjusting the transmittance of the filter. However, note that depending on the value of $\epsilon$ the contribution of $\left|E_{\rm sca}\right|^2$ may no longer be negligible. Thus, the value of $\epsilon$ affects which approximation that can be used when relating the scattering signal to particle properties.

A different approach to tune the signal-to-background ratio is to use a reflection rather than transmission geometry, as depicted in Figure \ref{fig:cobri}C\cite{lindfors2004detection}. This measurement geometry is in the literature typically denoted iSCAT (interferometric scattering)\cite{taylor2019interferometric}. When using a reflection geometry with transparent coverslips, a small portion of the illuminating light will be reflected back to the camera at the interface between the coverslip and the sample. This optical field constitutes the term $E_{\rm ill}$ in this geometry. Most of the light will be transmitted through the interface. This transmitted light produces the scattered field $E_{\rm sca}$ from particles in the sample. Denoting the reflectivity of the interface $\epsilon$, the recorded intensity is again identical to Eq. \eqref{eq:iscat}. In some works, these two approaches for background attenuation have been employed in parallel to achieve maximal sensitivity\cite{cole2017label,doi:10.1126/science.aar5839}. Moreover, by using two reflections from the top and the bottom of microfluidic channels, the interference between the two reflections can be used to control the phase and amplitude of the background signal\cite{olsen2024dual}. Thus, there are several ways of controlling the amplitude of the background signal, which in turn affects the relation between signal and polarizability. 

\subsection{Quantitative field microscopy}

Although interferometric scattering techniques have emerged as powerful techniques for particle characterization with low detection limits, only the real part of the scattered light is detected and quantified. Quantitative field microscopy, by contrast, achieves quantification of the full optical field, thereby providing a more complete characterization of the scattered light. Even though quantitative field microscopy historically has mostly been used to investigate samples such as live cells\cite{zangle2014live}, it has recently been shown that it is possible to detect particles down to single proteins\cite{thiele2024single}.

\begin{figure}[!ht]
\centering
\includegraphics[width=0.9\linewidth]{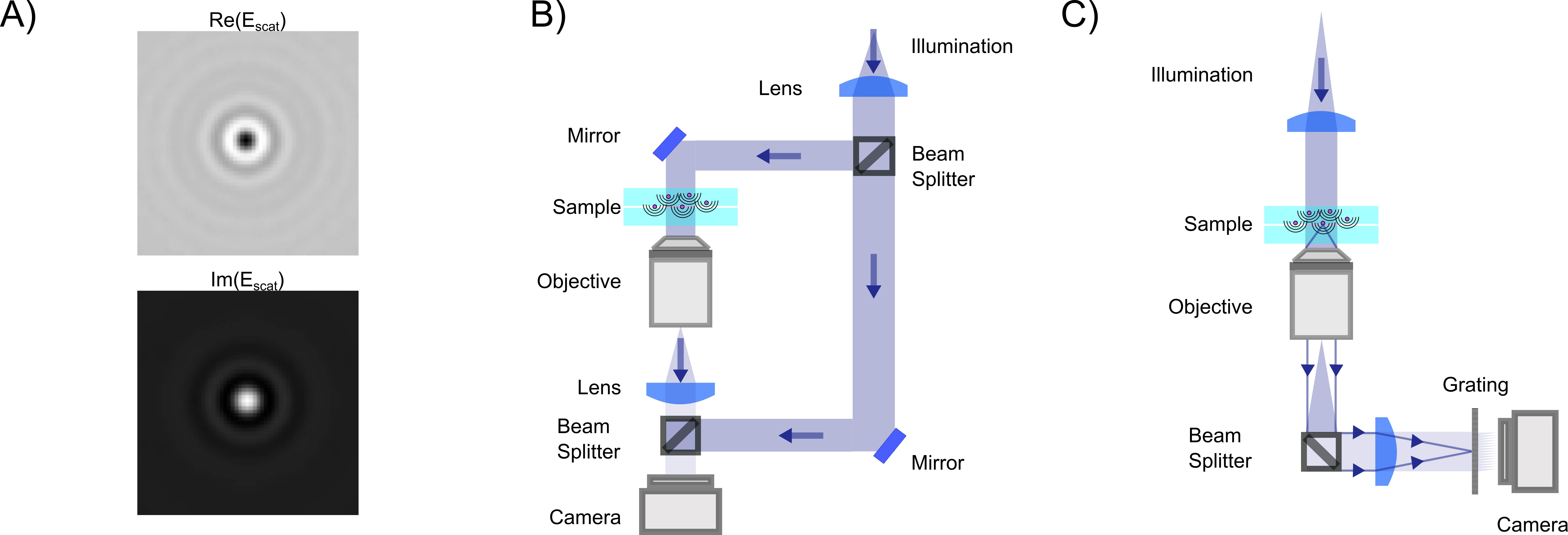}
\caption{\textbf{Quantitative field microscopy setups}: A) In quantitative field microscopy, both the real (upper) and imaginary parts (lower) of the scattered field are quantified, where the sign and amplitude of the real and imaginary part contain particle material information\cite{khadir2020full,midtvedt2019size}. B) Off-axis holography: the illumination light is split into two separate paths prior to the sample. The reference arm and the sample arm are recombined close to the camera at an angle. C) Quadriwave shearing microscopy (QWLS): the light is split using a diffraction grating after the sample. The reference light in this setup is a sheared version of the field impinging on the diffraction grating. }
\label{offAxis}
\end{figure}

In quantitative field microscopy, the light incident on the camera interferes with another reference field $E_{\rm ref}$ to achieve quantification of the real and imaginary parts of the scattered field itself (Figure\ref{offAxis}A). This can be achieved using several different microscopy configurations, both using a single image and multiple images to obtain the optical field\cite{kim2010principles,chaumet2024quantitative}. The trick to achieving single-image quantitative field microscopy is to let the reference field be incident on the camera at an angle $\theta_{\rm ref}$ such that $\sin\theta_{\rm ref}>3\sin\theta_{\rm max}/M$, where $M$ is the magnification of the optical system. From the interference with the reference field, it is then possible to quantify the optical field and its angular components (Box \ref{box:quantfield})\cite{cuche1999simultaneous}.

\begin{pabox}[label=box:quantfield]{Quantifying the optical field}
In quantitative field microscopy, a reference field is introduced at the camera plane, propagating at an angle $\theta_{\rm ref}$ with respect to the optical axis. Taking the reference field to propagate parallel to the $xz$-plane, the field at the camera plane can be represented as a plane wave as $E_{\rm ref}(x,y)=|E_{\rm ref}| e^{i\mathbf{k}_{\rm ref}\cdot \mathbf{x}}$, where $\mathbf{k}_{\rm ref}=k\sin \theta_{\rm ref} \hat{x}$. As a further simplifying assumption, we consider an illumination propagating along the optical axis. In this case, the intensity recorded by the camera reads
\begin{equation}
\begin{split}
    I_{\rm cam}^{\rm qf}(x,y) =& |E_{\rm ref}|^2 + |E_{\rm ill}|^2 + 2|E_{\rm ill}||E_{\rm ref}|\Re(e^{-ik\sin \theta_{\rm ref} x})\\ & +2|E_{\rm ill}|\Re(e^{i\varphi}E_{\rm sca}(x,y)^*) + 2|E_{\rm ref}|\Re(e^{ik\sin \theta_{\rm ref}x}E_{\rm sca}(x,y)^*)\\&  +|E_{\rm sca}|^2.
    \end{split}
    \label{eq:qffield1}
    \end{equation}
    
Now, let us investigate the plane wave decomposition (the Fourier transform) of this recorded intensity. Since typically $E_{\rm sca}\ll E_{\rm ill}$ and $E_{\rm sca}\ll E_{\rm ref}$, we keep only terms up to linear order in $E_{\rm sca}$. Further, we utilize that at the camera plane,
\begin{align}
    E_{\rm sca}(\mathbf{x}) = & \mathcal{F}^{-1}[\hat{E}_{\rm sca}(\mathbf{k}) \tilde{P}(\mathbf{k})] \\
    E_{\rm sca}(\mathbf{x})e^{-i\mathbf{k}_{\rm ref}\cdot \mathbf{x}} = &\mathcal{F}^{-1}[\hat{E}_{\rm sca}(\mathbf{k}-\mathbf{k}_{\rm ref}) \tilde{P}(\mathbf{k}-\mathbf{k}_{\rm ref})]
\end{align}
to arrive at
\begin{equation}
    \begin{split}
        \hat{I}_{\rm cam}^{\rm qf}(\mathbf{k})=& (|E_{\rm ref}|^2 + |E_{\rm ill}|^2)\delta(\mathbf{k}) + |E_{\rm ill}||E_{\rm ref}|(\delta(\mathbf{k}-\mathbf{k}_{\rm ref})+\delta(\mathbf{k}+\mathbf{k}_{\rm ref}))\\
        &+|E_{\rm ill}|\cos\varphi \left[\tilde{P}(\mathbf{k})\hat{E}_{\rm sca}(\mathbf{k}) + \tilde{P}(-\mathbf{k})\hat{E}_{\rm sca}(-\mathbf{k})\right] \\&+|E_{\rm ref}|\left[\tilde{P}(\mathbf{k}-\mathbf{k}_{\rm ref})\hat{E}_{\rm sca}(\mathbf{k}-\mathbf{k}_{\rm ref}) + \tilde{P}(\mathbf{k}_{\rm ref}+\mathbf{k})\hat{E}_{\rm sca}(\mathbf{k}_{\rm ref}+\mathbf{k})\right].
    \end{split}
\end{equation}

At this point, recall that the function $\tilde{P}$ has support only for arguments $|\mathbf{k}|<k\sin\theta_{\rm max}/M$. It is therefore possible to choose $\mathbf{k}_{\rm ref}$ such that $\tilde{P}(\mathbf{k})\tilde{P}(\mathbf{k}-\mathbf{k}_{\rm ref})=0$ for all values of $\mathbf{k}$. Specifically, this holds if $\sin\theta_{\rm ref}>2\sin\theta_{\rm max}/M$. In this case, one finds that
\begin{equation}
    \hat{I}_{\rm cam}^{\rm qf}(\mathbf{k}) \tilde{P}(\mathbf{k}-\mathbf{k}_{\rm ref}) = |E_{\rm ref}|\tilde{P}(\mathbf{k}-\mathbf{k}_{\rm ref})\hat{E}_{\rm sca}(\mathbf{k}-\mathbf{k}_{\rm ref}),
\end{equation}
from which the optical field can be reconstructed through an inverse Fourier transform. In this analysis, the term $|E_{\rm sca}|^2$ was neglected. To ensure that $\tilde{P}(\mathbf{k}-\mathbf{k}_{\rm ref})$ does not overlap with this term either, one can show that it is sufficient that $\sin\theta_{\rm ref}>3\sin\theta_{\rm max}/M$.

\end{pabox}

The reference field can be constructed via an external reference (referred to as off-axis holography, Figure \ref{offAxis}B) or by interference by the illuminating field itself, for instance, by introducing a grating at the camera plane (Figure \ref{offAxis}C)\cite{kim2010principles,chaumet2024quantitative}. The different implementations have some different strengths and weaknesses regarding complexity, noise, and stability. A detailed discussion about different microscopy configurations can be found in Ref. \cite{chaumet2024quantitative}.

Similar to interferometric scattering techniques, optical field measurements can also be combined with optical filters\cite{ortiz2022simultaneous,olsen2023label,olsen2024dual} and different illumination strategies\cite{saemisch2021one,thiele2024single} to improve the detection limit. In that case, the attenuation factor $\epsilon$ will affect the relation between the measured particle signal and particle properties.


\section{Scattering of light from nano- and microparticles}

In the preceding section, the optical field was taken to be a scalar quantity to simplify the equations. In reality, the optical field is a vector with components given by the polarization state of the light. The different polarization directions scatter light slightly differently, therefore complicating the description of the scattered light. Nonetheless, as we will show below, under certain circumstances, the scalar description of light still provides an accurate description of light scattering.

The scattering of light from nanoparticles can generally be described by solving Maxwell's equation for electromagnetism with the appropriate boundary conditions. In the special case of spherical particles, the exact solution to this problem was derived by Gustav Mie in 1908\cite{bohren2008absorption}. The field scattered by a homogeneous sphere is written as an infinite sum of special functions, which is numerically much faster to evaluate than explicitly solving Maxwell's equations. 

To gain insight into the behaviour of the scattered light and into how the light scattering is affected by particle properties, it is useful to consider limiting cases for which the Mie solution can be evaluated analytically. One such limit, which is of particular relevance for nanoparticles, is the \textit{Rayleigh limit}, valid for $kR\ll 1$, where $k=2\pi/\lambda$ and $R$ is a characteristic size of the particle. Let the illuminating light be a linearly polarized plane wave with amplitude $|E_{\rm ill}|$, propagating along the $z$-axis and defining the $x$-axis to lie along the polarization direction. Now, the scattered wave is described both by the angle $\theta$ relative to the propagation axis of the light and by the angle $\phi$ relative to the polarization axis. The scattered field will maintain the polarization direction of the illuminating field, and its plane wave decomposition is represented as
\begin{equation}
    \hat{E}_{\rm sca}(\theta,\phi)=|E_{\rm ill}|\frac{ik^3\alpha}{4\pi}\Lambda(\theta,\phi),
\end{equation}
where $\Lambda=\left(\cos^2\phi+\sin^2\phi\cos^2 \theta\right)$ for linearly polarized light, and $\alpha$ is the \textit{polarizability} of the particle, 
\begin{equation}
    \alpha=3V\frac{n_{\rm p}^2-n_{\rm m}^2}{n_{\rm p}^2+2n_{\rm m}^2}.
\end{equation}
In this expression, $V$ is the particle volume, $n_{\rm p}$ is the refractive index of the particle, and $n_{\rm m}$ is the refractive index of the surrounding medium. 

Another useful limit is the \textit{weakly scattering limit}, that is, for $kR|n_{\rm p}/n_{\rm m}-1|\ll 1$ and $|n_{\rm p}/n_{\rm m}-1|\ll 1$\cite{bohren2008absorption}. This is a slightly weaker condition than the Rayleigh condition, and the corresponding approximation is valid also for arbitrarily large scatterers as long as the refractive index difference compared to the surrounding medium is sufficiently small so that the inequality $kR|n_{\rm p}/n_{\rm m}-1|\ll 1$ still holds. This limit is particularly useful for biological nanoparticles, which typically obey these inequalities. In this case, the polarizability is typically approximated as
\begin{equation}
    \alpha\approx 2V\Delta n/n_{\rm m},
\end{equation}
where $\Delta n=n_{\rm p}-n_{\rm m}$. In the specific case of biological nanoparticles, this expression has a particular physical interpretation since it enables the treatment of the nanoparticle as a volume made up of biomolecules at a certain concentration. The refractive index of a solution of biomolecules increases approximately linearly with the mass concentration $C$ of molecules, as $n_{\rm p}=n_{\rm m} + (dn/dc)\cdot C$, where $(dn/dc)$ is called the specific refractive index increment, and is material specific. However, since most biomolecules contain similar elements at similar ratios (mostly carbon, hydrogen, oxygen, and nitrogen), the specific refractive index increments of different types of biomolecules are very similar\cite{zangle2014live}. Typical values range from $\sim 0.16$~ml/g for carbohydrates to $\sim 0.2$~ml/g for nucleic acids and proteins\cite{zangle2014live}. The polarizability times $n_{\rm m}$ thus evaluates to $n_{\rm m}\cdot\alpha\approx 2(dn/dc)C\cdot V=2(dn/dc)m$, where $m$ is the total mass of the biomolecules in the nanoparticle. Thus, in the case of weakly scattering particles, the polarizability is proportional to particle mass.

Further, the total scattered field can be calculated as the superposition of the field scattered by infinitesimal volume elements within the particle. The scattering from each such infinitesimal volume element is given by the Rayleigh scattering limit above. For weakly scattering particles, the optical field impinging on each such volume element can be approximated as equal to the incident optical field external to the particle.  The scattered field from an isotropic particle, evaluated outside of the particle, is then given by\cite{sorensen2001light}
\begin{equation}
    E_{\rm sca}(\theta,\phi) = ik^3|E_{\rm ill}| \Lambda(\theta,\phi) \int dr r^2 \Delta n(r)  \frac{\sin qr}{qr},
    \label{eq:RDG}
\end{equation}
where $q=2k\sin(\theta/2)$. This approximation is called the \emph{Rayleigh-Debye-Gans (RDG) approximation}.

The integral above describes the interference of the light scattered from different volume elements in the particle. This factor is denoted the \textit{form factor} and physically encodes the distribution of refractive index within the particle. The RDG field is often written in the following form,
\begin{equation}
    E_{\rm sca}(\theta,\phi) = ik^3 |E_{\rm ill}|\alpha \Lambda(\theta,\phi) f(\theta;R),
\end{equation}
where $f(\theta;R)$ is the form factor, which for an isotropic particle depends on its size $R$ and internal refractive index distribution.

Under uniform illumination, analytical solutions to the form factor within the RDG approximation can be attained for some specific geometries, such as spheres and core-shell spheres.
For a spherical particle, the form factor is \cite{wyatt1993light},
\begin{equation}
    f_{\rm sphere}(\theta)=\frac{3}{(qR)^3}\Big(\sin(qR)-qR\cos(qR)\Big),~
    \label{eq:sphere}
\end{equation}
and for an infinitesimal spherical shell, which can be used to approximate the signal from a lipid vesicle, the form factor is \cite{hallett1991vesicle}
\begin{equation}
    f_{\rm shell}(\theta)=\frac{\sin(qR)}{qR}.
    \label{eq:vesicle}
\end{equation}

\begin{figure}[!ht]
\centering
\includegraphics[width=.6\linewidth]{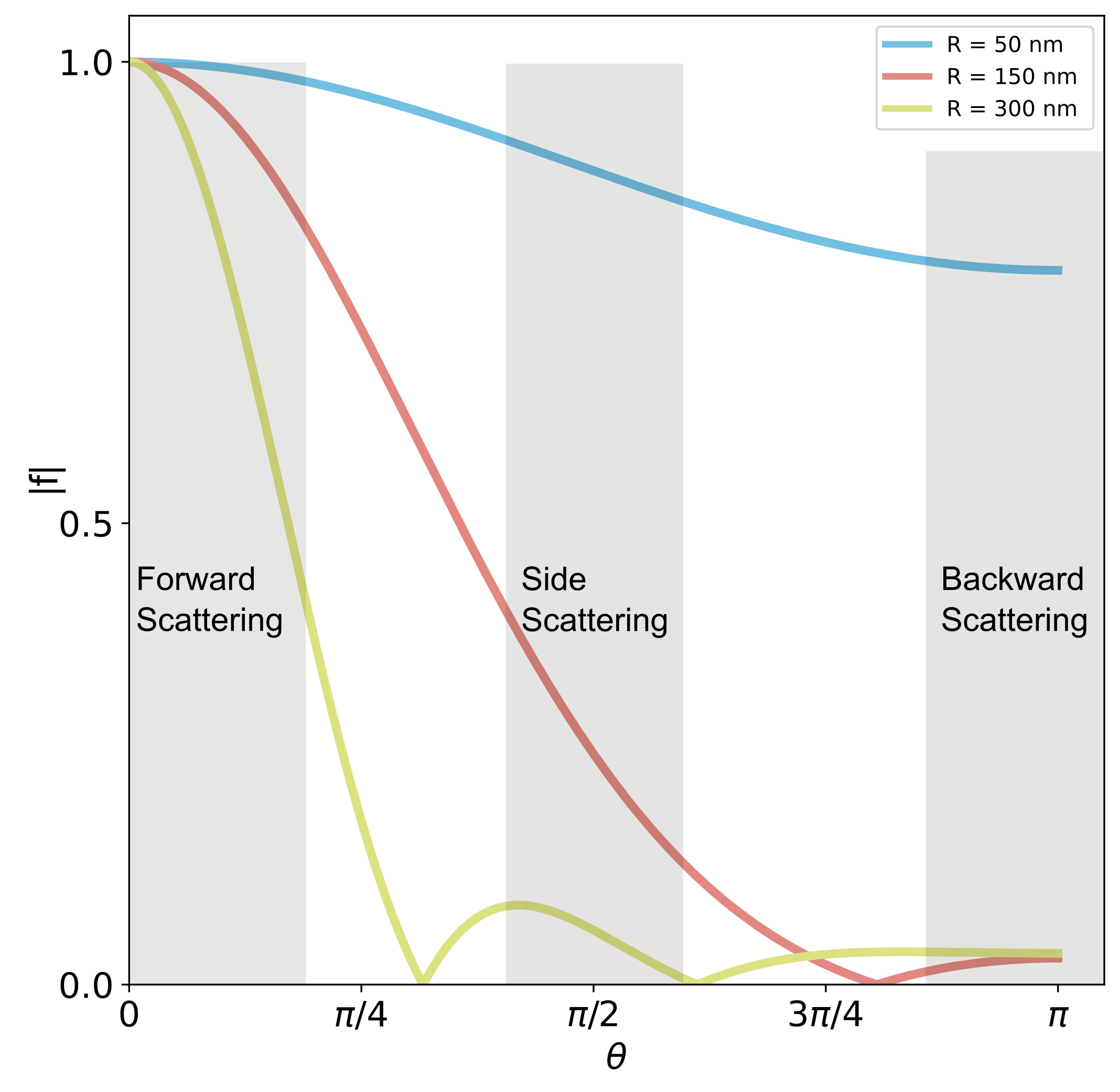}
\caption{\textbf{Form factors of homogeneous spheres.} The form factor depends strongly on the particle size. Small particles (blue line) scatter almost uniformly, while large particles (green line) scatter predominantly in the forward direction (small angles). In forward scattering, the contribution from the form factor is close to unity in most cases. In side scattering ($\theta_{\rm ill}\approx\pi/2$), the contribution is more pronounced, and in backward scattering ($\theta_{\rm ill}\approx \pi$), the form factor contribution is maximal. The form factors in this plot are calculated for homogeneous spheres in water illuminated with a wavelength of $\lambda$ = 532 nm. Shorterning the illumination wavelength will compress all scattering curves to smaller scattering angles. Conversely, for longer illumination wavelengths the scattering curves will be extended to the right in the figure above.}
\label{FormFactor_radius}
\end{figure}

Note that the form factor at $0$ degree scattering angle is identically equal to unity ($f(0)=1$) and that the form factor for all scattering angles $\theta>0$ is smaller than unity ($f(\theta>0)<1$). For this reason, the relation between the measured optical signal and particle size is different for different measurement geometries. This is highlighted in Figure \ref{FormFactor_radius}, in which the form factor of spheres of different sizes is shown as a function of the scattering angle. In accordance with the discussion above, the form factor is close to one for all particle sizes for transmission geometries, for which the scattering angle is close to $\theta=0$. For side-scattering (exemplified in Figure \ref{dark_field_filter}C) and backward scattering (exemplified in Figure \ref{fig:cobri}B), the form factor greatly influences the relationship between the measured optical signal and particle size. Specifically, the contribution from the form factor to the scattered light becomes appreciable for particles with $qR>1$. Recalling that $q=2k\sin(\theta/2)$, one has that the form factor contribution is appreciable if $R>(2k\sin(\theta_{\rm ill}/2))^{-1}$. For an illumination wavelength of $532$ nm when the particle is in water ($n_{\rm m}\approx 1.33$), this amounts to $R>45$ nm for side-scattering with $\theta_{\rm ill}=\pi/2$, and $R>30$ nm for backscattering with $\theta_{\rm ill}=\pi$.

\section{Relation between the signal measured in a scattering microscope and physical particle parameters}\label{sec:relation}
Now, let us use the mathematical framework developed in Sections 2 and 3 to investigate the relation between the scattered light from a nanoparticle and the optical field at the camera plane. The crucial insight is that the angular distribution of the scattered field from a nanoparticle described in Section 3 describes precisely the individual plane wave components used in Eq. (\ref{eq:bfpplanewaves}). Working within the RDG approximation and assuming that the illuminating light propagates along the optical axis, we write the scattered field as 
\begin{equation}
E_{\rm sca} (\theta,\phi) = i|E_{\rm ill}| k^3\alpha f(\theta)\Lambda(\theta,\phi),
\end{equation}
where $\alpha$ is the polarizability of the particle, $f(\theta)$ is the form factor, $|E_{\rm ill}|$ is the amplitude of the illuminating light. The polarization function $\Lambda(\theta,\phi)$ introduces an angular dependence of the scattered light that does not depend on the properties of the particles in the sample in the case of isotropic particles. In the process of signal calibration utilizing reference samples through darkfield and interferometric microscopy, this contribution will be effectively normalized away. Thus, to simplify the expressions, we will drop the polarization function $\Lambda(\theta,\phi)$ in the following sections of this tutorial. For a particle located in the focal plane of the objective (see Box \ref{box:outoffocus} for a description of the general case of a particle located away from the focal plane), the scattered field at the back focal plane is then given by
\begin{equation}
    E_{\rm bfp}(\rho) =ik^3|E_{\rm ill}| \alpha  \int_{0}^{\theta_{\rm max}} \cos\theta \sin\theta f(\theta) \hat{P}(k(\rho-f\sin\theta)) d\theta
\end{equation}
and the scattered field at the camera plane is given by
\begin{equation}
    E_{\rm sca}(\rho) = i|E_{\rm ill}| k^3\alpha \int_{0}^{\theta_{\rm max}} \cos\theta \sin\theta f(\theta) J_0(k\rho\sin\theta)d\theta.
    \label{eq:RDGscattered}
\end{equation}

\begin{pabox}[label=box:outoffocus]{Effect of having particle located away from the focal plane}
Equation \eqref{eq:RDGscattered} is valid for a scatterer located at the focal plane of the objective. In an experiment, this condition is not necessarily fulfilled. If the particle is located a distance $z$ from the focal plane, the individual plane wave components of the scattered field must additionally be propagated to the focal plane. Since each of the components describes a plane wave propagating at an angle $\theta$ with respect to the optical axis, the angular components propagated a distance $z$ along the optical axis are given by
\begin{equation}
    \hat{E}(\theta,z) = \hat{E}(\theta,0)e^{ikz\cos\theta}.
    \label{eq:FieldProp}
\end{equation}
Thus, the scattered field from a nanoparticle located a distance $z$ from the focal plane will, at the camera plane, reads
\begin{equation}
    E_{\rm sca}(\rho) = i|E_{\rm ill}|k^3 \alpha  \int_{0}^{\theta_{\rm max}} \cos\theta \sin\theta f(\theta) J_0(k\rho\sin\theta)e^{ikz\cos\theta}\ d\theta.
\end{equation}

\end{pabox}
We will now use this expression, in combination with the definition of the RDG form factor Eq. \eqref{eq:RDG}, to investigate how the signal measured in a scattering microscope is related to the physical parameters of the scattering objects.

\subsection{Darkfield microscopy}
\begin{figure}[!ht]
\centering
\includegraphics[width=\linewidth]{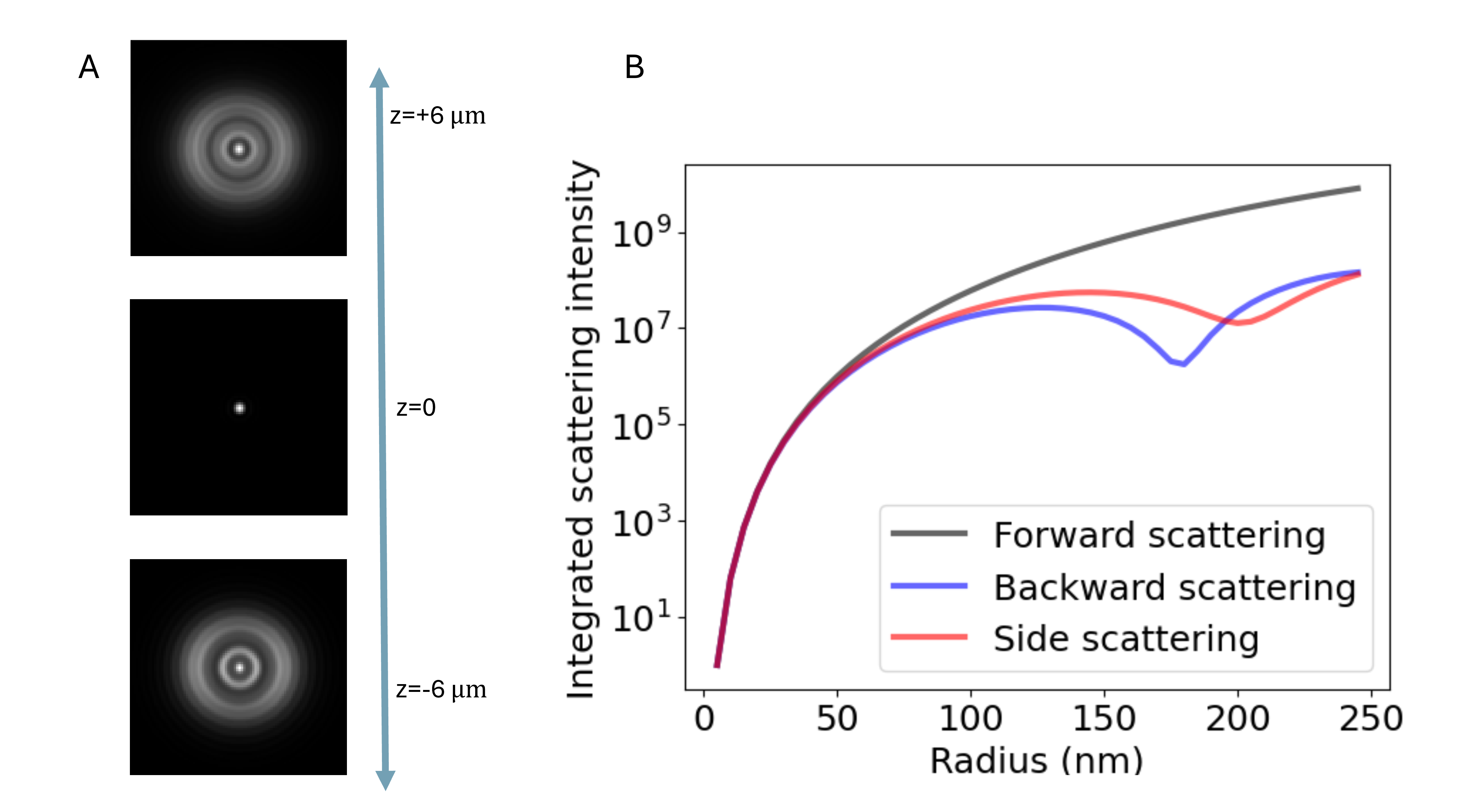}
\caption{\textbf{Scattered intensity in darkfield imaging} A) Calculated scattering patterns of nanoparticles (radius 20 nm) suspended in water and measured in darkfield microscopy with $\theta_{\rm ill}=0$ and wavelength 532 nm at different values of the defocus $z$. The field of view of each particle image is $10\times 10$ microns. B) The integrated intensity of scattering patterns measured in darkfield microscopy as a function of particle size for three illumination angles. For $\theta_{\rm ill}=0$, the integrated intensity scales with the square of the volume throughout the sizes included in this calculation. For $\theta_{\rm ill}\neq 0$, distinct minima appear corresponding to the minima of the form factor.}
\label{fig:DFintensity}
\end{figure}

In the case of darkfield microscopy, the intensity measured by the camera from a particle located at the front focal plane of the objective and illuminated by a plane wave propagating along the optical axis is given by
\begin{equation}
    I_{\rm camera}(\rho) =|E_{\rm ill}|^2 k^6|\alpha|^2  \left|\int_{0}^{\theta_{\rm max}} \cos\theta \sin\theta f(\theta) J_0(k\rho\sin\theta)e^{ikz\cos\theta}\ d\theta\right|^2.
    \label{eq:DFscattering}
\end{equation}
Some examples of scattering patterns obtained from Eq. \eqref{eq:DFscattering} at different defocus values $z$ are shown in Figure \ref{fig:DFintensity}A.

To perform particle characterization, one needs to reduce the measured scattering to a set of values representing some physical properties of the particle. The most common way of achieving this is by characterizing the integrated intensity of a scattering pattern, which in the case of darkfield microscopy is proportional to the square of the polarizability. In the case of a particle illuminated by a plane wave with $\theta_{\rm ill}=0$, one obtains for an isotropic scatterer (see Box \ref{box:intdfield} for a derivation of this result)
\begin{equation}
    \int I_{\rm camera} dA = 2\pi |E_{\rm ill}|^2 k^4|\alpha|^2 \int_{0}^{\theta_{\rm max}}\cos\theta \sin\theta f(\theta)^2 d\theta.
    \label{Eq:DFIntensity}
\end{equation}
Similar expressions for cases in which the illuminating field is not parallel to the optical axis can be derived by adjusting the limits of integration in the above expression. Furthermore, for small enough scatterers such that $f(\theta)\approx 1$, the integrated darkfield intensity is proportional to the square of the polarizability. For larger particles ($kR>1$), the contribution to the measured signal from the form factor of the scatterer will depend on their size and morphology\cite{hannestad2020single}. 

\begin{pabox}[label=box:intdfield]{Integrated darkfield intensity}
In order to calculate the integrated scattering intensity in darkfield microscopy, it is useful to start with the expression
\begin{equation}
    I_{\rm cam}(\mathbf{x}) = |E_{\rm sca}(\mathbf{x})|^2.
\end{equation}

Integrating this over the entire detector surface, one has 
\begin{equation}
    \int I_{\rm cam}(\mathbf{x}) d\mathbf{x} = \int |E_{\rm sca}(\mathbf{x})|^2 d\mathbf{x}.
\end{equation}

Now, we invoke Parseval's theorem, stating that
\begin{equation}
    \int |f(x)|^2  dx = \int |\hat{f}(k)|^2 dk,
\end{equation}

where $f(x)$ and $\hat{f}(k)$ are Fourier transform pairs. One therefore has
\begin{equation}
    \int I_{\rm cam}(\mathbf{x}) d\mathbf{x} = \int |\hat{E}_{\rm sca}(\mathbf{k})|^2 d\mathbf{k},
\end{equation}

from which Eq. \eqref{Eq:DFIntensity} follows after transformation into polar coordinates and switching integration variable from the wave vector $|\mathbf{k}|$ to angle $\theta$. 

\end{pabox}
The approaches to achieve darkfield illumination discussed in Section 2 give rise to slightly different contributions from the form factor. For darkfield imaging through spatial blocking, the illumination angle $\theta_{\rm ill}=0$ in Eq. \eqref{Eq:DFIntensity}. In the case of darkfield illumination by oblique illumination, one instead has $\theta_{\rm ill}\neq 0$. The final case of evanescent wave scattering is slightly more complicated. In this case, the optical field is propagating along the interface in which total internal reflection occurs. Therefore, the illumination angle is $\theta_{\rm ill}=\pi/2$. However, the form factor is slightly different since the optical field decays exponentially in the medium of the scatterers. The appropriate correction was derived in the supporting information of Ref.~\cite{hannestad2020single}. In Figure \ref{fig:DFintensity}B, the integrated scattered intensity is shown as a function of particle size for a fixed refractive index for the specific cases $\theta_{\rm ill}=0$, $\theta_{\rm ill}=\pi/2$ and $\theta_{\rm ill}=\pi$. As discussed in Section 3, the larger the $\theta_{\rm ill}$, the smaller the size region for which there is a unique relation between scattering signal and size for a known particle refractive index.


\subsection{Interferometric scattering techniques}
\begin{figure}[!ht]
\centering
\includegraphics[width=\linewidth]{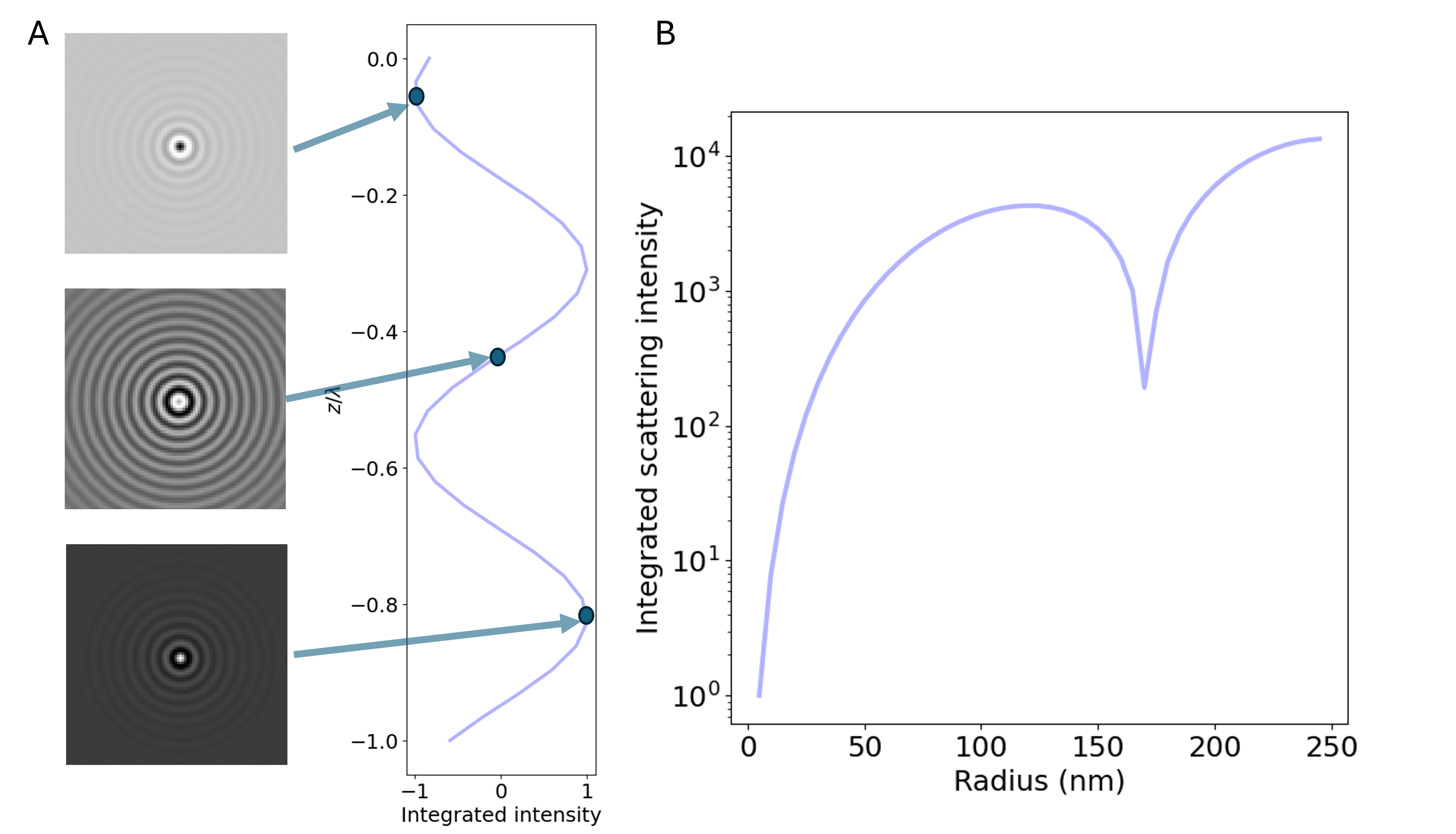}
\caption{\textbf{Scattered intensity in interferometric imaging} A) Calculated scattering patterns of nanoparticles (radius 20 nm) suspended in water and measured in interferometric microscopy in reflection geometry ($\theta_{\rm ill}=\pi$) with illumination wavelength 532 nm at different values of the defocus $z$, indicated in the plot to the right of the scattering patterns. The integrated intensity of the scattering patterns varies sinusoidally with the defocus $z$ due to the phase shift $\varphi$. The field of view of each particle image is $10\times 10$ microns. B) The integrated intensity of scattering patterns measured in interferometric microscopy with $\theta_{\rm ill}=\pi$ as a function of particle size.}
\label{fig:iscatintensity}
\end{figure}
In the case of interferometric scattering techniques, the term of interest in Eq. \eqref{Eq:intscat} is the final interferometric term. Taking the illuminating field to be a plane wave propagating along the optical axis, the integrated recorded intensity is given by (see Box \ref{box:intintint} for a derivation)
\begin{equation}
    \int \left(\frac{I_{\rm camera}}{I_0}-1\right) d\mathbf{x} = (2/\epsilon)k \Re(\alpha e^{i\Delta\varphi}) f(\theta_{\rm ill}),
    \label{eq:iscat_normalised}
\end{equation}
where $\Delta\varphi=\varphi-\varphi(z)$ is the relative phase difference between the background field and the scatterer, $\epsilon$ is the attenuation factor, and $\varphi(z)$ describes the depth-dependent relative phase shift of the particle relevant for reflection geometries\cite{dong2021fundamental}. Further, we have defined $I_0 = \epsilon^2 |E_{\rm ill}|^2$, which is experimentally estimated by evaluating the intensity recorded by the camera in locations without particles present.
Note that the right-hand side of this final expression does not depend on the intensity of the illuminating light. In other words, the illuminating light does not need to be separately quantified to perform particle characterization in interferometric scattering approaches. However, since $\epsilon$ may have a spatial dependence or vary between different surfaces\cite{becker2023quantitative,olsen2024dual}, background corrections are still sometimes needed. Also, in contrast to darkfield approaches, the recorded intensity is now directly proportional to the real part of the polarizability when $\Delta\varphi=0$.

\begin{pabox}[label=box:intintint]{Integrated interferometric intensity}
    To evaluate the integrated signal from a particle in interferometric microscopy, we write the illumination field at the camera plane as 
\begin{equation}
    \tilde{E}_{\rm ill} = \epsilon|E_{\rm ill}| e^{i(\varphi\pm kz)},
\end{equation}
where $\epsilon$ is the attenuation coefficient of the illuminating light, $\varphi$ is the phase shift of the illuminating field compared to the scattered field, and $z$ is the distance from the scatterer to the focal plane. The $\pm$ reflects the fact that the light propagates in opposite directions compared to the original propagation direction for reflection and transmission geometries. The $+$ sign is relevant for reflection geometries, and the $-$ sign for transmission geometries. To the lowest order in the scattered field, the recorded intensity, in the RDG approximation and assuming an isotropic scatterer, is
\begin{equation}
    I_{\rm cam}(\rho) =\epsilon^2 |E_{\rm ill}|^2+ 2\epsilon|E_{\rm ill}|^2 \Re\bigl[\alpha k^3 e^{i(\varphi\pm kz)} \hspace{-5pt} \int_{0}^{\theta_{\rm max}}\hspace{-5pt} \cos\theta \sin\theta f(|\theta-\theta_{\rm ill}|) J_0(k\rho\sin\theta)e^{ikz\cos\theta} d\theta\bigr],
    \label{eq:iscatscatter}
\end{equation}
where $\theta_{\rm ill}=0$ for interferometric imaging with transmission geometry, and $\theta_{\rm ill}=\pi$ for interferometric imaging with reflection geometry. The first term in this equation represents the intensity recorded by the camera in the absence of scatterers. Subtracting and dividing by this value, one obtains for the integrated signal of a particle
\begin{equation}
\begin{split}
    \int \frac{\delta I}{\epsilon^2|E_{\rm ill}|^2} d\mathbf{x} = (2/\epsilon) k^3 \Re{\bigl[\alpha} e^{i(\varphi\pm kz)}\hspace{-5pt}\int \hspace{-5pt} \int_{0}^{\theta_{\rm max}}\hspace{-5pt} \cos\theta \sin\theta f(|\theta-\theta_{\rm ill}|) \rho J_0(k\rho\sin\theta)e^{ikz\cos\theta} d\rho d\theta\bigr],
    \end{split}
\end{equation}
where $\delta I = I_{\rm cam}-\epsilon^2 |E_{\rm ill}|^2$.
To evaluate this, recall that the integral over $\theta$ is a Fourier transform in disguise. Utilizing that 
\begin{equation}
    \int f(x) dx = \hat{f}(0),
\end{equation}
where $f$ and $\hat{f}$ are Fourier transform pairs, one immediately finds
\begin{equation}
    \int \frac{\delta I}{\epsilon^2|E_{\rm ill}|^2} d\mathbf{x} = (2/\epsilon)k\Re{[\alpha} e^{i(\varphi+[1\pm 1]kz)}] f(\theta_{\rm ill}).
\end{equation}
\end{pabox}
To exemplify the position dependence of the interferometric scattering signal, in Figure \ref{fig:iscatintensity}A scattering patterns calculated according to equation \eqref{eq:iscatscatter} in the case for interferometric backscattering. Notice how the central lobe changes sign as the defocus changes by only a fraction of a wavelength. The integrated signal of the scattering patterns shows a sinusoidal dependence on the defocus $z$ due to the relative phase difference $\varphi(z)$ between the scattered light from the particle and the light reflected at the coverslip. For this reason, the iSCAT signal is either quantified in the same plane for all particles, as on a coverslip or a specific depth plane\cite{doi:10.1126/science.aar5839,kashkanova2021precision}, or the images are transformed using a neural network to remove the depth depth dependence\cite{olsen2024dual}. When that is done accurately, the size dependence of the integrated signal follows Figure \ref{fig:iscatintensity}B, which shows Eq. \eqref{eq:iscat_normalised} when $\Delta \varphi=0$.


\subsection{Quantitative field imaging}
\begin{figure}[!ht]
\centering
\includegraphics[width=\linewidth]{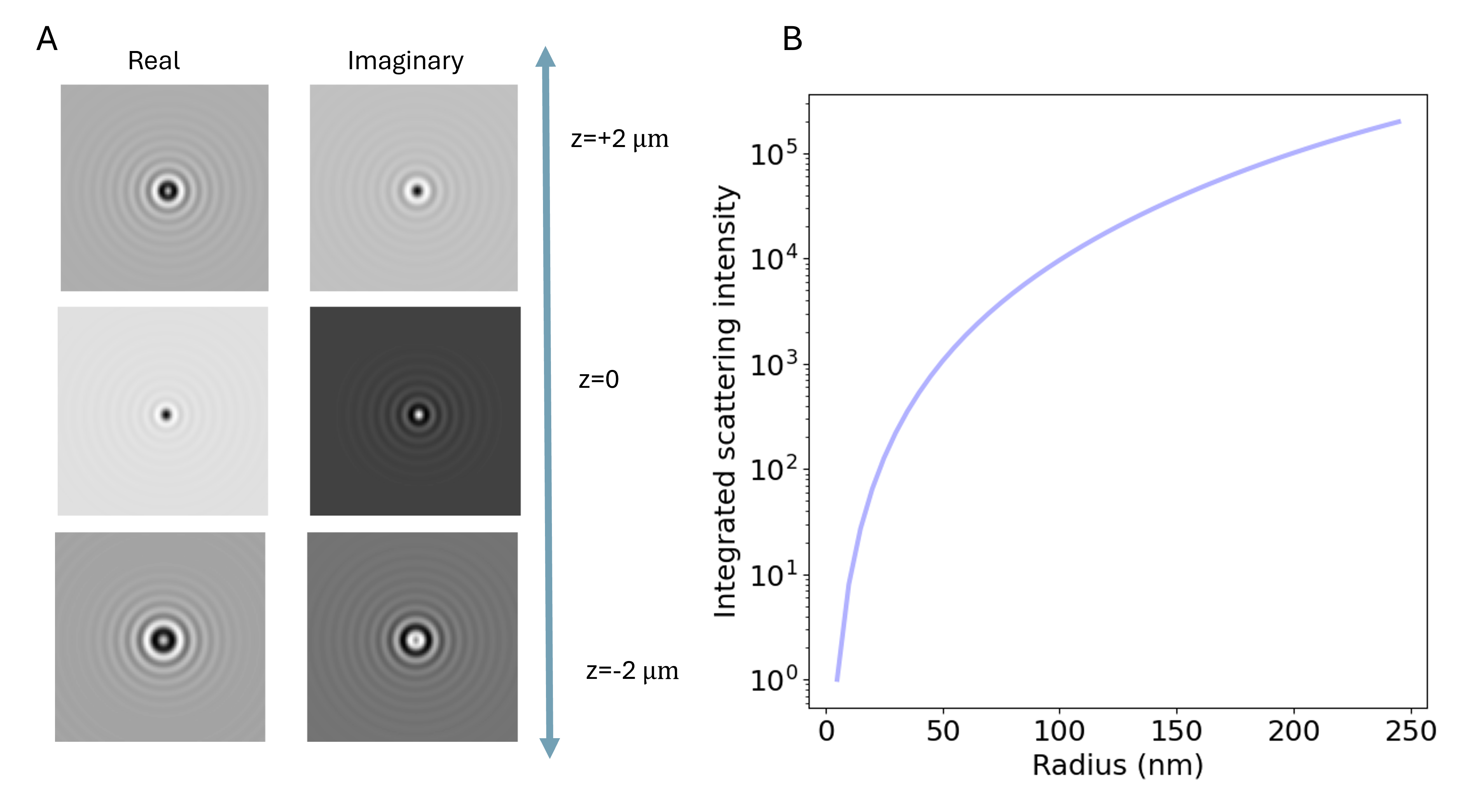}
\caption{\textbf{Scattered intensity in quantitative field imaging} A) Calculated scattering patterns (real and imaginary parts) of nanoparticles (radius 20 nm) suspended in water and measured in quantitative field imaging with illumination wavelength 532 nm at different values of the defocus $z$. The field of view of each particle image is $10\times 10$ microns. B) When the particle is in focus, the integrated imaginary part of scattering patterns is measured in quantitative field microscopy scales with particle volume. Note that for quantitative field imaging, the image can be re-propagated after recording. This makes it possible to refocus the detections individually.}
\label{fig:QFintensity}
\end{figure}

In quantitative field imaging, the recorded quantity is the scattered field itself. In this case, one has that\cite{khadir2020full}
\begin{equation}
    \int E_{\rm sca}(\mathbf{x})/|E_{\rm ill}| d\mathbf{x} = i k\alpha f(\theta_{\rm ill})=i k\alpha,
    \label{eq:intfieldqf}
\end{equation}
where the last equality is valid for $\theta_{\rm ill}=0$, which is the most common choice for quantitative field methods. This is similar to Eq. \eqref{eq:iscat_normalised} with $\varphi=0$, except for the fact that the polarizability is now allowed to be complex-valued. Note that it is the integrated imaginary part of the optical field that is proportional to the real part of the particle polarizability. In Figure \ref{fig:QFintensity}A, the real and imaginary parts of the scattering patterns measured in quantitative field imaging are shown for different defocus values $z$. Importantly, following Eq. \eqref{eq:FieldProp} the optical field signal can be re-propagated after recording the image. This, in turn, enables quantification of the signal of focused scattering patterns even though they are measured out of focus, which reduces the sensitivity of the particle characterization to noise and out-of-focus effects\cite{khadir2020full}. The integrated imaginary part of the signal is proportional to particle polarizability and hence particle volume (Figure \ref{fig:QFintensity}B) as predicted from Eq. \eqref{eq:intfieldqf}. 

Furthermore, utilizing Eq. \eqref{eq:FTransform} one can rewrite the scattered field at the camera plane in terms of the angular components of the form factor directly (Box \ref{box:qfformfact}). This shows how quantitative field imaging contains information about particle polarizability and the particle form factor itself, which, if quantified, can be related to particle size\cite{wyatt1998submicrometer}.

\begin{pabox}[label=box:qfformfact]{Quantifying the optical form factor in quantitative field imaging}

The scattered field at the camera plane is related to the optical form factor as
\begin{equation}
    E_{\rm sca}(\rho) = ik^3 |E_{\rm ill}| \alpha \int \cos\theta \sin\theta f(\theta) J_0(k\rho \sin\theta)e^{ik z \cos\theta} d\theta.
\end{equation}

Applying the transform Eq. \eqref{eq:FTransform}, one finds that
\begin{equation}
    i k \alpha |E_{\rm ill}| f(\theta) e^{ik z \cos\theta} =  \int \rho E_{\rm sca}(\rho) J_0(k\rho \sin\theta) d\rho.
\end{equation}

Taking the absolute value of both sides and utilizing that $|e^{ix}|=1$, one finds

\begin{equation}
    k |\alpha| |E_{\rm ill}| |f(\theta)|=  \left|\int \rho E_{\rm sca}(\rho) J_0(k\rho \sin\theta) d\rho\right|.
\end{equation}
Finally, since $k |\alpha| |E_{\rm ill}| = |\int E_{\rm sca} d\mathbf{x}|$ from Eq. \eqref{eq:intfieldqf}, we have that
\begin{equation}
    |f(\theta)|=  \frac{|\int \rho E_{\rm sca}(\rho) J_0(k\rho \sin\theta) d\rho|}{|\int E_{\rm sca} d\mathbf{x}|}.
    \label{eq:FormFacFromField}
\end{equation}
\end{pabox}


\section{A toolbox for analyzing scattering microscopy data}

Quantitative analysis of scattering microscopy data consists of two fundamental steps, namely \textit{particle detection} and \textit{signal characterization}\cite{midtvedt2022single}. The aim of this section is to provide an easy-to-use toolbox to perform these tasks for the three types of scattering microscopy geometries that are discussed in this tutorial. If the particles are moving during the experiment, a third step, \textit{detection linking}, is required to form particle traces\cite{manzo2015review}, allowing the particle motion to be related to particle properties such as the hydrodynamic radius as previously described in several review articles\cite{chenouard2014objective,bian2016111}. Since the ability to track the motion of particles is generic for optical microscopy methods, the focus of this tutorial is the information contained in the optical signal, where the hydrodynamic radius will complement the optical signal in the case of freely suspended particles. Appended to this tutorial are Jupyter notebooks containing code for performing particle detection and characterization in the three scattering modalities considered here (darkfield imaging, interferometric imaging, and quantitative field imaging). In the following two sections, the content of the notebooks will be briefly explained.

\subsection{Particle detection}
The particle detection task is essentially recognizing scattering patterns in the presence of noise. Traditionally, particle detection has been performed by algorithmic approaches, in which a predefined set of image filters are applied to the microscopy images, followed by a thresholding operation to identify particles\cite{rishi2015particle}. In the past decade, deep learning approaches to particle detection have become increasingly popular, showcasing more accurate detection in particular under low signal-to-noise conditions\cite{midtvedt2021quantitative,midtvedt2022single}.

There are, in general, three sources of noise contributing to the noise level in scattering microscopy images: (\textbf{1}) \emph{shot noise}, arising from the finite number of photons detected in each camera pixel, (\textbf{2}) \emph{read noise}, which is intrinsic to the camera, and (\textbf{3}) \emph{speckle noise}, due to coherent reflections and scattering along the beam path of the illuminating light\cite{meiniel2018denoising}. 

The first two noise sources are common to all types of microscopy and have the property of being spatiotemporally independent: the noise at pixel $i$ at time $t_0$ is independent of the noise at pixel $j$ at time $t_1$. This particular property means that the noise from these sources can be reduced by averaging the signal over time and/or across multiple pixels. 
Speckle noise is special for optical microscopy and originates from the interference between the different optical plane waves of the illumination. Speckle noise is characterized by the fact that noise at neighboring pixels is correlated, where the amplitude and temporal stability of the speckle depends on the light source and experimental setup\cite{bianco2018strategies}. Important for image analysis, this noise has a spatial correlation that is similar to the spatial correlation of the nanoparticle scattering signal, and in the absence of mechanical vibrations in the system, it can be considered static. Thus, neither temporal nor spatial averaging helps to reduce the effect of this particular noise term. This noise source is primarily important for interferometric approaches, where the particle contrast is determined relative to the illuminating optical field\cite{bianco2018strategies}.

To improve the detection limit, both background subtraction and signal averaging are commonly used, where the approach depends on whether the particles are immobilized or freely diffusing. For freely diffusing particles, the background can be subtracted by averaging adjacent frames to the current frame as the background features are static and the particle of interest is moving\cite{midtvedt2019size,kashkanova2021precision,olsen2024dual}. When working with such background subtraction, it is important that the background frames are chosen so that the particle signal is not subtracted. Moreover, for freely diffusing particles, it is difficult to improve the detection by averaging frames, as the particles are at different positions in each frame. A special case is when looking at particles binding to a surface, where both a rolling background subtraction and frame averaging can be applied\cite{cole2017label}. 
For immobile particles, frame averaging can be used to improve the detection limit\cite{baffou2012thermal,cole2017label}, where the background subtraction is typically done by using images prior to the particle binding\cite{sjoberg2021time}. 

After background subtraction, the particles are often readily detectable using several different methods. 
In the notebooks, particle detection using both deep learning analysis (specifically LodeSTAR \cite{midtvedt2022single}) and an algorithmic approach (radial variance transform \cite{kashkanova2021precision}) is demonstrated on calculated images of scattering patterns for the different modalities. An example of the particle detection step using LodeSTAR is shown in Figure \ref{fig:iscatnb}A for the case of interferometric imaging with reflection geometry. To handle the change of signal for different particle depth positions, the network is trained on a range of different z-positions, which enables accurate detection of the particles in the image.


\subsection{Signal characterization}\label{sec:sigChar}
After having detected a particle, the next step in nanoparticle characterization is to utilize the image of the scattering pattern to extract information about the particle itself.

Common to all scattering microscopy approaches is that the integral of the particle signal is related to particle polarizability (which, for biological nanoparticles, is proportional to their mass). However, depending on the measurement geometry, when relating the scattering to mass, the effect of the optical form factor needs to be compensated (Figure \ref{FormFactor_radius}). As a rule of thumb, for particle sizes $R<(2k\sin\theta_{\rm ill})^{-1}$, the form factor can be approximated as $f(\theta)\approx 1$ within the angles collected by the objective, in which case the integrated intensity is directly related to particle polarizability. For particles larger than this, the integrated intensity must be complemented with independent measurements of size and/or polarizability to perform quantitative characterization, as exemplified in \cite{olsen2024dual,kashkanova2023label}.

\begin{figure}[!ht]
\centering
\includegraphics[width=.6\linewidth]{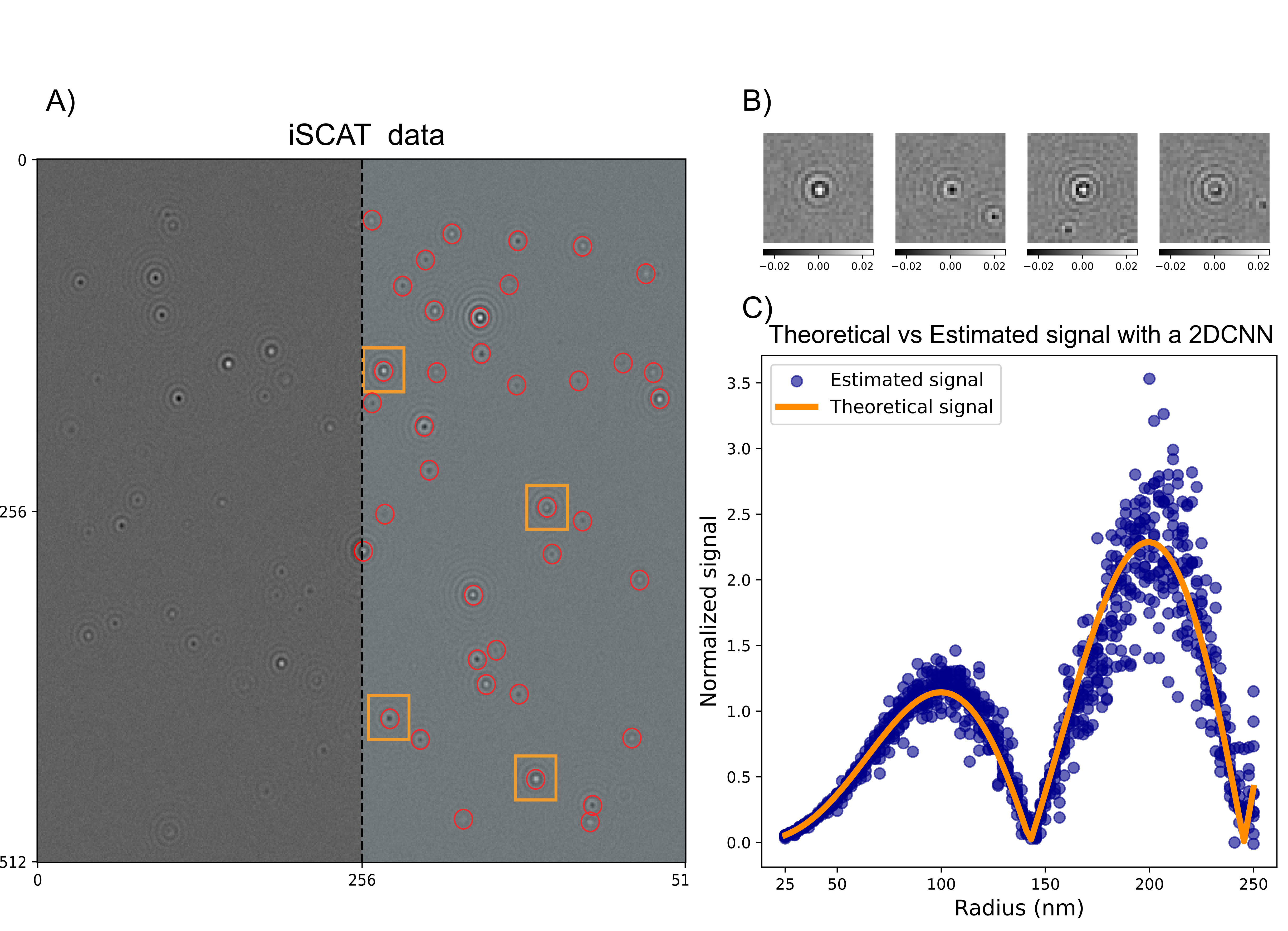}
\caption{\textbf{Particle detection and characterization in interferometric microscopy} A) An example iSCAT microscopy image containing multiple scatterers. This image was simulated using the notebooks appended to this tutorial. On the right side of the image, particle detections made by the LodeSTAR algorithm, trained using the notebooks appended to this tutorial, are overlaid on the image. The particle detections are intentionally left out on the left half of the image to not obscure the appearance of the scattering data. B) Crops of the scattering patterns of individual particles detected in the frame in A). C) Characterization of the integrated scattering intensity of individual particle crops like the ones shown in B). The characterization is performed by a convolutional neural network, trained using the notebooks appended to the tutorial.}
\label{fig:iscatnb}
\end{figure}

In practice, the task in signal characterization is to estimate the integrated particle signal in the presence of noise. Directly summing up all camera pixels is not a good approach in practice since, most commonly, tens or hundreds of particles are present in the field of view of the camera at the same time. The most common approach to signal characterization is, therefore, to crop out a small region around each detected particle (Figure \ref{fig:iscatnb}B), and fit some kind of function to this limited view of a particle. As a specific example in the case of iSCAT, in \cite{kashkanova2023label} the particles were first localized using the radial variance transform, where the particle detection with the maximum positive contrast estimated by Gaussian fitting was used to estimate the particle signal.


Another approach that has gained increasing attention is to utilize deep learning enhanced analysis techniques to not only detect the particle but also estimate the signal. In the appended notebooks, we provide code to train and apply a convolutional neural network to estimate the integrated signal strength of scattering patterns. In Figure \ref{fig:iscatnb}C, this step is shown for interferometric imaging in reflection geometry, where the signal estimation follows what is expected from theory. In the appended notebooks, code is also provided to perform this step for darkfield imaging as well as for quantitative field imaging.

Going beyond particle polarizability, it is, in some instances, possible to quantify also the particle form factor directly from the optical signal using deep learning image analysis by utilizing the fact that the form factor is encoded in the angular components of the scattered field as measured in quantitative field imaging (Eq. \eqref{eq:FormFacFromField}). This approach was utilized in Ref.~\cite{midtvedt2021fast} to estimate particle size and refractive index directly from scattering patterns measured in off-axis holographic images. Specifically, particle size (or, more accurately, the radius of gyration of a particle) is related to its scattering form factor as when $Rq\ll 1$\cite{debye1947molecular}
\begin{equation}
    R_{\rm g}^2 \approx 3q^{-2}(1-f(\theta)^2).
\end{equation}
Thus, the task of particle sizing directly from quantitative field images amounts to estimating a curvature in the optical form factor from noisy images.

\section{Considerations when designing measurement geometry}



From the treatment above, it is clear that the different approaches to scattering microscopy have different quantitative power when it comes to particle characterization. 

The first consideration that one should make when designing a scattering-based characterization experiment is the level of detail required in the characterization to answer the scientific questions at hand. In some cases, it may be sufficient to detect and track the motion of particles rather than accurately quantify the particle signal. In this case, darkfield techniques have the advantage that the data is relatively easily analyzed since the particles appear bright against a dark background. For this reason, darkfield imaging is one of the standard techniques for tracking suspended nanoparticles\cite{van2014refractive,kim2019validation}. 

\begin{figure}[!ht]
\centering
\includegraphics[width=\linewidth]{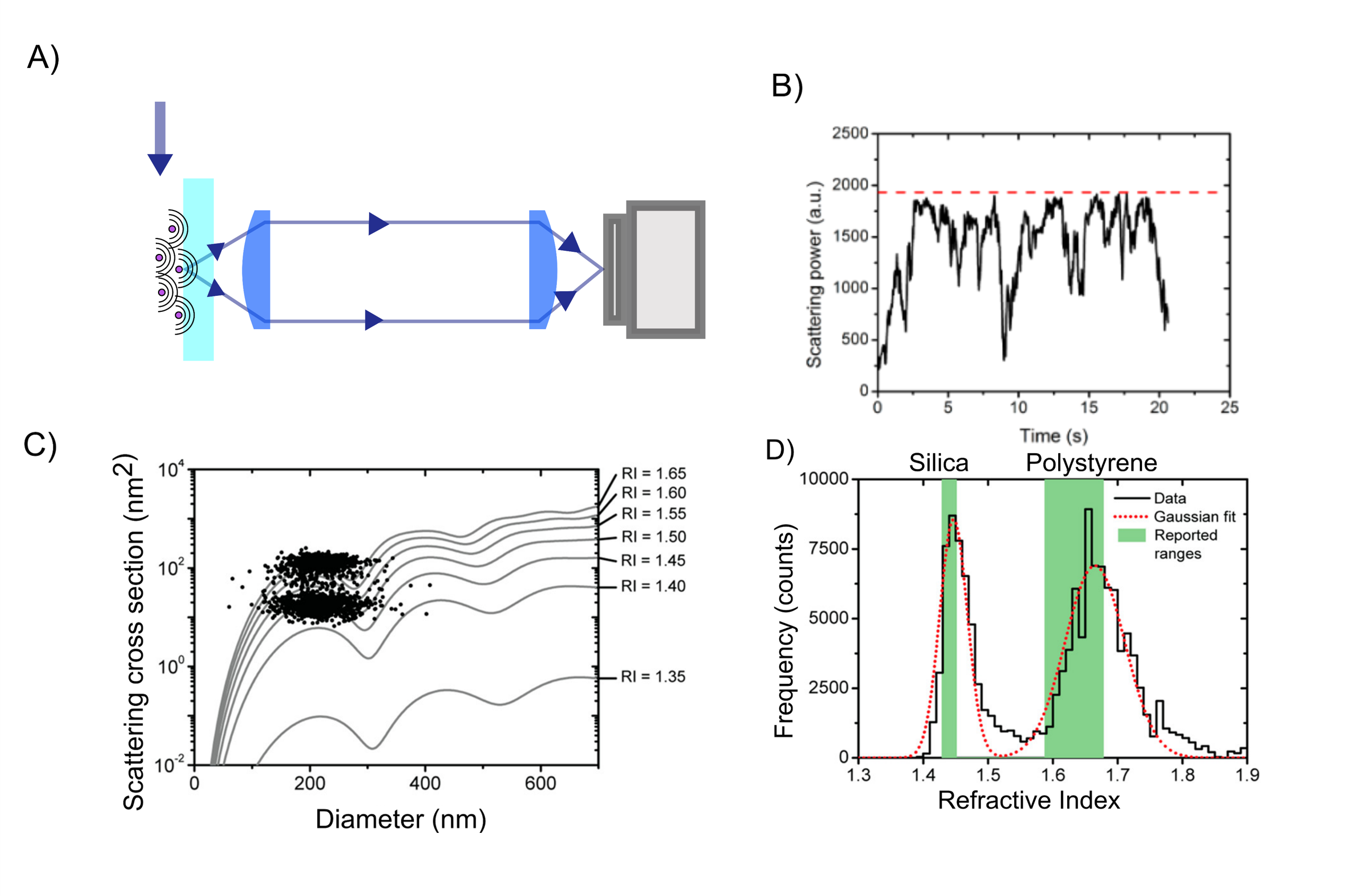}
\caption{\textbf{Particle characterization using darkfield imaging with oblique illumination} A) In \cite{van2014refractive} darkfield imaging with oblique illumination was used for quantitative characterization of suspended nanoparticles. B) The particle signal was estimated by tracking the motion of nanoparticles and estimating the integrated signal at each time point in a particle trace. The maximum value of the integrated signal was used as a proxy for the signal strength. C) Using both signal quantification and particle tracking over time enabled quantification of both hydrodynamic size and the scattering cross-section, here for a sample of silica beads and a sample of polystyrene beads. D) By combining the particle size and scattering cross section the particle refractive index was also estimated for the silica and polystyrene beads. Figure reprinted with permission from American Chemical Society (Copyright 2014).}
\label{vanderpol}
\end{figure}

However, particle characterization based on the optical signal using darkfield techniques is comparatively challenging since relating particle signal to polarizability requires accurate calibration. In particular, since the particle contrast in darkfield techniques is proportional to the local light intensity at the particle position, a proper calibration procedure would require mapping out the illumination intensity throughout the entire field of view, which is technically challenging. In \cite{van2014refractive}, particle characterization was demonstrated using darkfield imaging with oblique illumination (Figure \ref{vanderpol}A) of a sample freely diffusing in a macroscopic volume. The challenge of quantifying the scattered signal in such conditions, in particular under a non-uniform illumination, was overcome by utilizing the maximum value of the measured scattering signal of each particle trace as a proxy for the particle scattering, in combination with careful calibration (Figure \ref{vanderpol}B). This enabled the quantification of hydrodynamic size as well as the scattering cross-section of suspended polystyrene and silica beads (Figure \ref{vanderpol}C), which was also converted into the estimate for the refractive index of these particles (Figure \ref{vanderpol}D). 

\begin{figure}[!ht]
\centering
\includegraphics[width=0.5\linewidth]{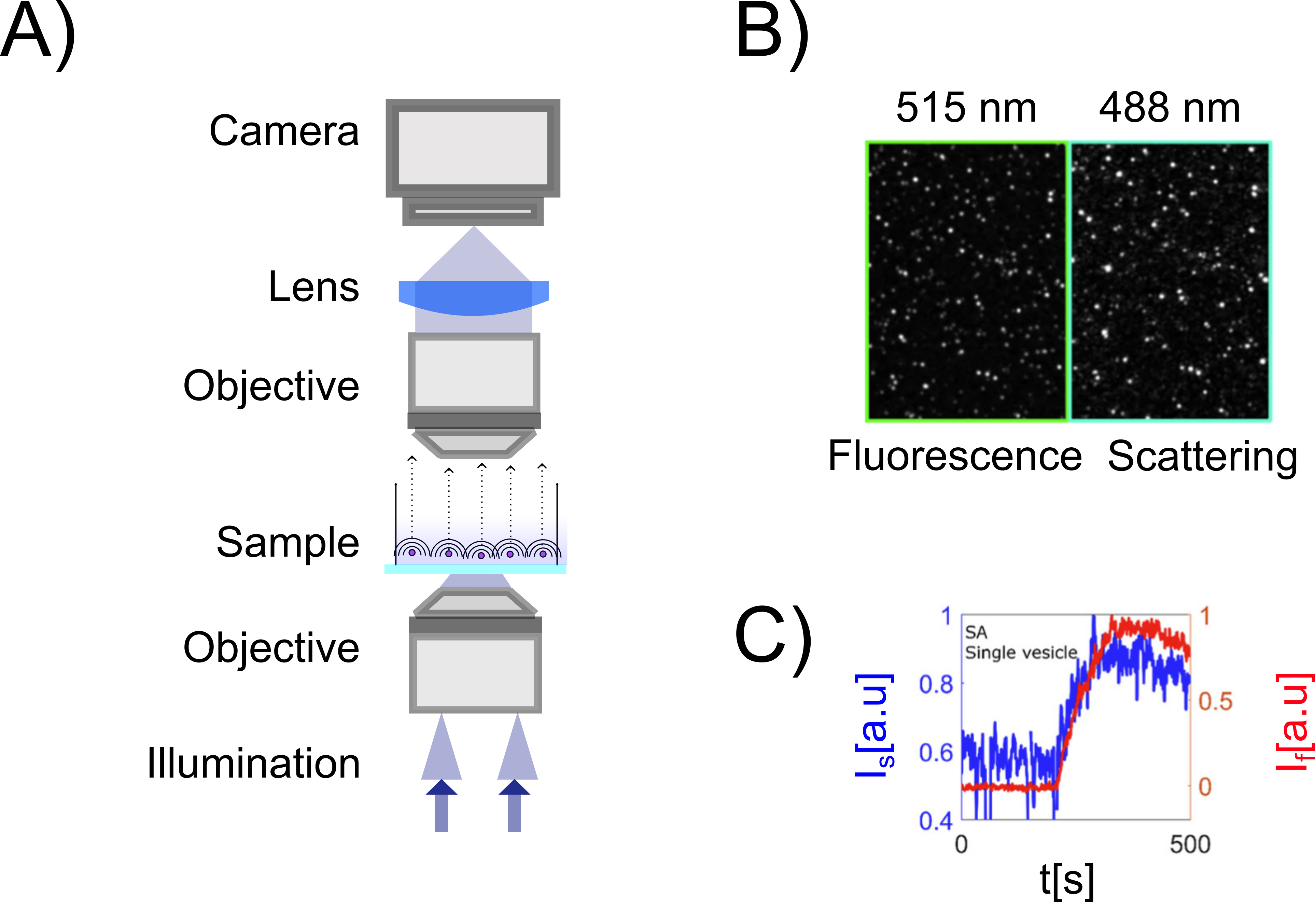}
\caption{\textbf{Characterization of surface-bound particles using evanescent illumination} A) In Ref. \cite{sjoberg2021time} evanescent illumination was used to study protein binding to lipid vesicles adsorbed to a surface. B) Using a dichromatic mirror, the scattering and fluorescence signals can be simultaneously recorded. C) The protein adsorption event could be resolved by monitoring the integrated fluorescence and scattering intensities as a function of time on the single particle level. Figure reprinted with permission under the CC-BY license.}
\label{evanescentcharac}
\end{figure}

Another measurement consideration is whether particle dynamics information is critical or not. To follow the same particle over time it needs to be confined, which can be achieved by, for example, tethering the particle to a surface\cite{sjoberg2021time}. In particular, by using evanescent illumination (Figure \ref{evanescentcharac}A), only particles that are adsorbed or very close to the surface will be illuminated and scatter light. In \cite{sjoberg2021time}, evanescent illumination was used to study protein adsorption to lipid vesicles adsorbed to a surface. Both fluorescence and scattering signal were measured simultaneously (Figure \ref{evanescentcharac} B), enabling time-resolved monitoring of the adsorbed protein mass to individual vesicles and correlating the scattering signal to fluorescence signal (Figure \ref{evanescentcharac}C).

A third measurement consideration is whether the particle signal must be accurately related to particle properties such as mass.
Interferometric scattering approaches have the advantage compared to darkfield techniques in that the particle contrast is measured relative to the local illumination intensity so that the particle estimate is insensitive to changes in the illuminating light intensity. This enables accurate quantification of the scattering signal that can be related to particle properties in a precise manner\cite{kashkanova2021precision,nanochannel2022,doi:10.1126/science.aar5839}. Nonetheless, measuring the scattering from well-characterized calibration particles is still necessary to calibrate the attenuation factor $\epsilon$ and the relative phase difference $\varphi$. For suspended particles, which diffuse in three dimensions, the particles will quickly explore a volume sufficiently large to cover all possible values of $\varphi$, rendering calibration of this phase unnecessary in this case. 

\begin{figure}[!ht]
\centering
\includegraphics[width=\linewidth]{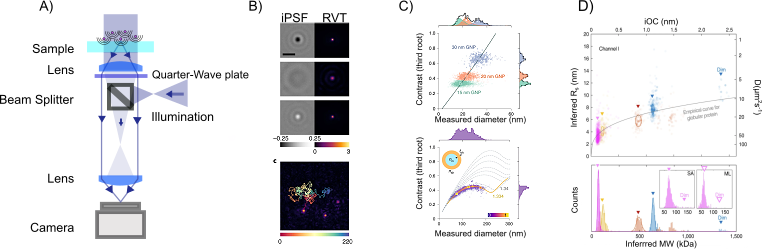}
\caption{\textbf{Particle characterization using iSCAT and nanochannel scattering microscopy (NSM).} A-C: In Ref.~\cite{kashkanova2023label}, particle characterization of suspended nanoparticles using iSCAT (A) was demonstrated through quantitative analysis of the scattered light in combination with particle tracking (B). The iSCAT contrast was shown to be proportional to particle volume for small particles (C, upper row), as anticipated from Section \ref{sec:relation}, and the deviation from this scaling was used to estimate the internal refractive index of extracellular vesicles (B, lower row). D) By utilizing the interference between a nanochannel and particles residing within the nanochannel, in \cite{nanochannel2022}, characterization of polarizability (proportional to mass) as well as the hydrodynamic radius of individual biomolecules was demonstrated. Figure reprinted with permission under the CC-BY 4.0 license.}
\label{iSCATparticlecharact}
\end{figure}

Interferometric methods enables accurate particle characterisation both on a surface\cite{doi:10.1126/science.aar5839} and when in solution\cite{kashkanova2023label,nanochannel2022}.
In Ref.~\cite{kashkanova2023label}, signal quantification in combination with particle tracking using iSCAT (Figure \ref{iSCATparticlecharact}A) was used to determine the size and refractive index of suspended nanoparticles (Figure \ref{iSCATparticlecharact} B-C). Moreover, by analyzing the particle-size scaling, they could obtain structural information about suspended liposomes. To investigate even smaller suspended particles, in Ref.~\cite{nanochannel2022}, they used the relative scattering between a nanochannel and the particle to characterize the suspended size and mass of individual biomolecules possible (Figure \ref{iSCATparticlecharact}D).

\begin{figure}[!ht]
\centering
\includegraphics[width=\linewidth]{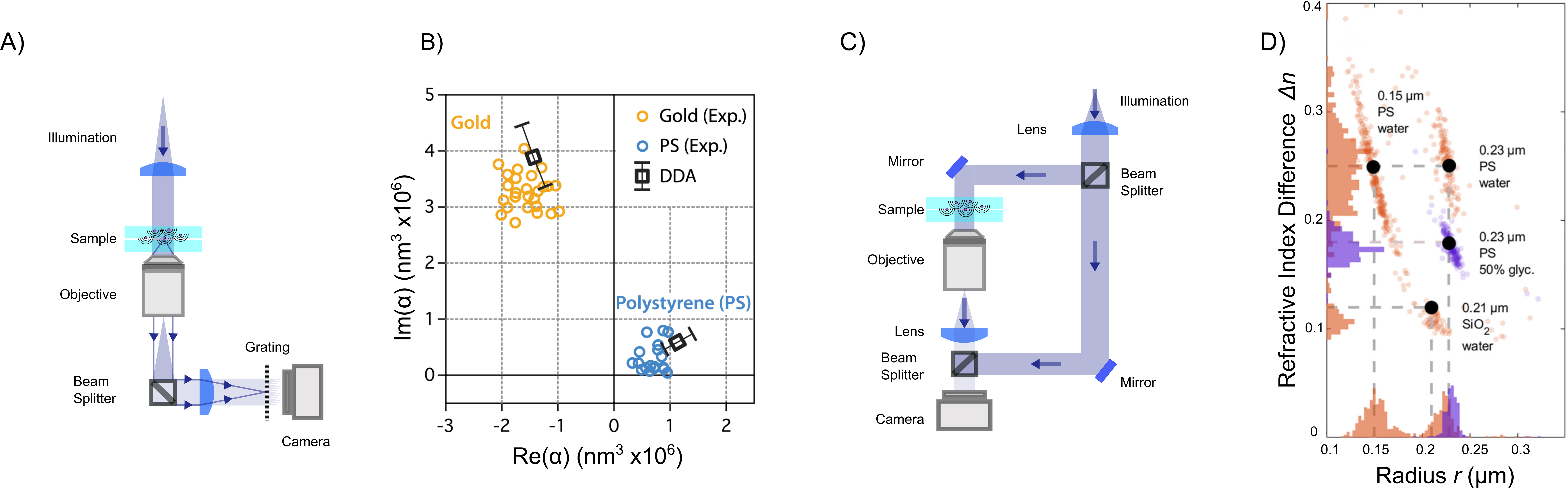}
\caption{\textbf{Particle characterization using quantitative field imaging.} A-B) Using quadriwave shear interferometric imaging, in Ref.~\cite{khadir2020full}, it was demonstrated that gold and dielectric nanoparticles can be distinguished based on their complex-valued polarizabilities. C-D) In Ref.~\cite{midtvedt2021fast}, it was demonstrated that deep learning enhanced analysis of scattering patterns recorded in off-axis holography is capable of quantifying the size and refractive index of suspended subwavelength particles. Figure reprinted with permission under the CC-BY 4.0 license.}
\label{qfparticlecharact}
\end{figure}

A fourth measurement consideration is whether more detailed material information is needed in the case of heterogeneous samples.
Quantitative field imaging provides the most rich optical signal that can, in turn, be used for the detailed characterization of nanoparticles. For instance, the complex-valued scattered field contains information about both the real and imaginary parts of the particle polarizability and refractive index (Figure \ref{qfparticlecharact}A-B)\cite{nguyen2023label,khadir2020full}. In Ref.~\cite{khadir2020full}, this was used to distinguish between gold nanoparticles and polystyrene particles directly from the optical signal. Similarly, in Ref.~\cite{midtvedt2019size}, the sign of the phase signal was used to differentiate between nanobubbles and dielectric particles in the same sample. Furthermore, in Ref.~\cite{midtvedt2021fast}, the complex-valued signal was used to determine the size and refractive index of nanobeads and fractal aggregates directly from the scattered light, without invoking particle tracking (Figure \ref{qfparticlecharact}C-D)\cite{midtvedt2021fast}. Thus, the quantitative field provides more detailed information about the particle material of the measured particles than darkfield and interferometric imaging. 

One major drawback of quantitative field microscopy techniques compared to other optical techniques has been its detection limit. However, it was recently shown that by combining evanescent field imaging with an external reference, it is possible to measure the optical field from single proteins with mass below 100 kDa when binding to a surface\cite{thiele2024single}.

This leads to another critical consideration when deciding on a measurement technique, namely particle size. The lowest reported detection limit is 9~kDa biomolecules measured using iSCAT (corresponding to a diameter of approximately 3 nm)\cite{dahmardeh2023self}. As a comparison, the reported detection limit for darkfield microscopy is around 5 nm diameter for non-metallic particles\cite{priest2021scattering}, iSCAT for suspended nanoparticles has a reported detection limit of around 40~nm \cite{kashkanova2021precision}, and quantitative field measurements for suspended nanoparticles has a reported detection limit of around 70~nm\cite{nguyen2023label}. In addition to the detection limit, as described in Section \ref{sec:sigChar}, transmission methods and non-transmission methods have different relationships between particle signal and polarizability. For particle mass determination, it is beneficial to use a measurement geometry for which the form factor contribution to the scattered light can be neglected. In practice, this implies that the illumination angle ideally should be related to the typical size $R$ of scatterers in the sample as $\sin (\theta_{\rm ill}/2)<(2kR)^{-1}$. 

Generally speaking, the more parameters that can be quantified at the single particle level, the more likely it is that the particles of interest can be distinguished from other particles in the sample. For instance, if the particles of interest are the strongest scattering particles, then any technique that permits quantifying polarizability is sufficient. If the particles scatter light to a similar extent but have a different material composition, then particle refractive index may be the relevant parameter to characterize. Finally, if the particles of interest differ from other particles in how the mass is distributed within the particle, then some technique capable of resolving the form factor is preferable. If these physical parameters are insufficient for distinguishing the particles of interest from other particles in the sample, it is also possible to augment scattering microscopy techniques with fluorescent imaging to achieve better specificity. 

\section{Conclusions and future opportunities}

In this tutorial, we have given an overview of the particle information from darkfield, interferometric scattering, and quantitative field microscopy measurements.
Over the past decades, the single particle detection limit has significantly improved for all different optical imaging methods. Looking ahead, there are still opportunities and challenges for scattering-based microscopy characterization beyond detection.
This includes multiparametric characterization of individual particles in terms of particle material, size, and shape, particularly for particles in complex environments such as inside cells.

\subsection{Optical fingerprinting}

A fundamental limitation in scattering-based particle characterization is that the detection events are nonspecific, as mentioned in the previous section. A challenge in the field of scattering-based nanoparticle characterization is to identify optical fingerprints, which enable distinguishing and characterizing subpopulations in heterogeneous samples without introducing labels.

One such fingerprinting feature, the integrated scattering amplitude, has been discussed at length in this tutorial. This feature is proportional to particle polarizability and can be used to distinguish subpopulations. Another experimentally measurable feature is the hydrodynamic radius, estimated through particle tracking\cite{van2014refractive,kashkanova2021precision,nanochannel2022}. 

By combining the information-rich scattering patterns of nanoparticles with deep learning-enhanced analysis techniques that go beyond quantifying the integrated scattering signal, we anticipate that more examples of fingerprinting features will be added to this list, enabling more precise population discrimination and characterization. To give a few examples: 
\begin{itemize}
    \item In principle, quantitative field imaging techniques can quantify the scattering form factor across all scattering angles collected by the objective. Utilizing this, it is possible to discriminate particle subpopulations based on their morphology.
    \item In biological systems, many processes are driven by weak interactions. In a scattering microscope, such interactions will manifest themselves as temporal fluctuations in particle properties\cite{midtvedt2021fast}. Such temporal fluctuations can also be used as a fingerprinting feature.
    \item Interferometric scattering patterns are, just as the field measured in quantitative field microscopy, related to the form factor evaluated over the scattering angles captured by the objective. This information can likely be decoded using deep learning enhanced analysis techniques, enabling precise sizing of very small objects.
    \item Shape information of anisotropic particles is also encoded in the scattering patterns. For suspended anisotropic nanoparticles, the scattered light reaching the camera will temporally fluctuate as the nanoparticle undergoes rotational diffusion. If the exposure time is much shorter than the characteristic time of rotational diffusion, these fluctuations can be resolved in measurement and be related to particle anisotropy\cite{guerra2019single}. Anisotropy is also encoded in the dependence of light scattering on polarization. This has been used to measure anisotropy of surface-bound particles\cite{vala2021quantitative}.
    \item Finally, the information about the scattering form factor within an image captured in a scattering microscope is limited by the scattering angles collected by the objective. Thus, different scattering approaches carry complementary information about the scattering form factor. By combining measurement modalities, it is possible to obtain a more complete mapping of the scattering form factor of individual nanoparticles. This was recently demonstrated by combining quantitative field imaging and iSCAT to quantify particle size in unknown sample media\cite{olsen2024dual}, and we anticipate that the same idea will be applied using other combinations of scattering techniques as well.
    
\end{itemize}

The primary obstacles to achieving such fingerprinting lie in the noise level of interferometric systems, obscuring parts of the scattering signal, and an imperfect characterization of the pupil function $P(\theta)$. In the treatment presented in this tutorial, this has been assumed to be perfectly characterized. In practice, this function is affected by aberrations in the optical system and is difficult to characterize perfectly. Furthermore, in this tutorial, all particles have been assumed to be located directly above the central line of the objective. In a real experiment, the position of the particle with respect to the center of the objective will affect the scattering angles that reach the objective. Thus, the image of a particle will be slightly affected by its position\cite{denis2015fast}. These effects were characterized and accounted for in \cite{midtvedt2021fast} using quantitative field microscopy, where calibration particles were used to obtain information about the occurring point spread function, but performing such characterization for other scattering microscopy geometries has not been demonstrated. Nonetheless, considering the fast improvement of the detection limits of scattering-based microscopy \cite{thiele2024single,doi:10.1126/science.aar5839} we anticipate that deep learning enhanced analysis of scattering patterns will enable precise nanoparticle characterization over a wide range of particle sizes and shapes.

\subsection{Characterizing particles in complex environments}

Most single particle characterization methods operate in known environments at a controlled particle concentration. Operating in unknown and crowed environments adds several challenges, including the unknown viscosity hindering estimation of the hydrodynamic radius and the overlapping scattering pattern hindering single particle tracking. For example, the cellular interior is both crowded and inhomogeneous, which makes the analysis of scattering patterns more complicated. 

All-optical fingerprinting of nanoparticles would alleviate the need for diffusion measurements in particle characterization, opening up the possibility of performing quantitative particle characterization in environments that were previously limited to qualitative particle analysis. For highly scattering particles, deep learning approaches can be used to identify and characterize the particles inside cells\cite{midtvedt2021quantitative}. For less strongly scattering particles, another approach to resolve the particles is to perform confocal scattering microscopy\cite{kuppers2023confocal,hsiao2022spinning}, in which only a small volume of the sample is illuminated at a time, thereby minimizing the effect of such unwanted light scattering. The depth selectivity of confocal microscopy significantly reduces the background scattering from other particles, allowing tracking of individual viruses on a cell\cite{kuppers2023confocal}. Thus, the combination of confocal scattering microscopy and deep learning image analysis will likely significantly extend the quantitative possibilities of label-free optical particle characterization. 

\section{Supporting information}
The code for detecting and characterizing nanoparticles can be found at\\ \url{https://github.com/softmatterlab/OpticalCharacterizationNanoparticles}





%


\section{Conflicts of interest}
DM and EO owns shares in a company that holds IP for quantification of particles in off-axis holographic imaging (HOLTRA). FH owns shares in the company Nanolyze, developing optical waveguide chips for surface sensitive scattering. DM and GV own shares in the company IFLAI, providing AI-based tools to analyze microscopy data.


\section{Acknowledgments}

This research was funded by the Swedish research council, grant number 2019-05071, the Knut and Alice Wallenberg Foundation grant number 2019-0577, Chalmers Area of Advance Nano, the Horizon Europe ERC Consolidator Grant MAPEI (Grant No. 101001267), and the Knut and Alice Wallenberg Foundation (Grant No. 2019.0079).

\section{Full contact information }
Erik Olsén -  Department of Physics, Chalmers University of Technology, SE-41296 Gothenburg, Sweden. ORCID: 0000-0002-4002-0917\\
Berenice García Rodríguez -  Department of Physics, University of Gothenburg, SE-41296 Gothenburg, Sweden. ORCID: 0009-0009-3572-0482\\
Fredrik Skärberg -  Department of Physics, University of Gothenburg, SE-41296 Gothenburg, Sweden\\
Giovanni Volpe -  Department of Physics, University of Gothenburg, SE-41296 Gothenburg, Sweden. ORCID: 0000-0001-5057-1846\\
Fredrik H{\"o}{\"o}k -  Department of Physics, Chalmers University of Technology, SE-41296 Gothenburg, Sweden. ORCID: 0000-0003-1994-5015\\
Daniel Sundås Midtvedt -  Department of Physics, University of Gothenburg, SE-41296 Gothenburg, Sweden. ORCID: 0000-0003-4132-4629

\section{Present Address Notes}
$^{\dagger}$ Erik Olsén is currently at Michael Smith Laboratories, University of British Columbia, Vancouver, BC V6T 1Z4
Canada

\bibliographystyle{ieeetr} 
\bibliography{references}


\end{document}